\documentclass[10pt]{article}

\usepackage[margin=20mm]{geometry}
\usepackage{setspace}
\usepackage{amsmath, amssymb, amsfonts}
\usepackage{graphicx}
\usepackage{booktabs}
\usepackage{hyperref}
\usepackage{enumitem}
\usepackage{caption}
\usepackage{subcaption}
\usepackage{float}
\usepackage{tabularx}
\usepackage{longtable}
\usepackage{siunitx}
\usepackage{xcolor}
\usepackage[section]{placeins}
\usepackage{array} 
\usepackage{mathtools}
\usepackage{parskip}
\usepackage{tikz}
\usetikzlibrary{arrows.meta,positioning,fit,calc,shapes.geometric}
\usepackage{pdfpages}

\usepackage{newtxtext,newtxmath}

\usepackage{csquotes} 
\usepackage[backend=biber,style=apa]{biblatex}
\DeclareLanguageMapping{american}{american-apa}
\title{Forward-Looking Correlation Forecasting for Statistical Arbitrage:\\
A Temporal--Heterogeneous Graph Neural Network (THGNN) Approach}
\author{}
\date{\today}

\begin{document}
\title{Forecasting Equity Correlations with Hybrid Transformer Graph Neural Network}

\author{
Jack Fanshawe\textsuperscript{1}\thanks{Corresponding author: jf3791@columbia.edu},
Rumi Masih\textsuperscript{1},
Alexander Cameron\textsuperscript{1}
}

\date{01/07/2025}

\maketitle
\begin{center}
\textsuperscript{1} Business School, Faculty of Business, Economics and Law,  
The University of Queensland, Australia
\end{center}

\maketitle

\section{Abstract}
This paper studies forward-looking stock-stock correlation forecasting for S\&P 500 constituents and evaluates whether learned correlation forecasts can improve graph-based clustering used in basket trading strategies. We cast 10-day ahead correlation prediction in Fisher-z space and train a Temporal-Heterogeneous Graph Neural Network (THGNN) to predict residual deviations from a rolling historical baseline. The architecture combines a Transformer-based temporal encoder, which captures non-stationary, complex, temporal dependencies, with an edge-aware graph attention network that propagates cross-asset information over the equity network. Inputs span daily returns, technicals, sector structure, previous correlations, and macro signals, enabling regime-aware forecasts and attention-based feature and neighbor importance to provide interpretability. Out-of-sample results from 2019-2024 show that the proposed model meaningfully reduces correlation forecasting error relative to rolling-window estimates. When integrated into a graph-based clustering framework, forward-looking correlations produce adaptable and economically meaningfully baskets, particularly during periods of market stress. These findings suggest that improvements in correlation forecasts translate into meaningful gains during portfolio construction tasks.

\section{Introduction}

Statistical arbitrage (StatArb) is one of the most actively studied and applied domains in quantitative finance, drawing on decades of research into asset co-movements and mean reversion strategies. Early approaches relied on static sector classifications such as, the Fama-French 48 industry groups (Fama \& French, 1997), with more recent strategies utilizing similarity metrics – correlation, cointegration, or distance – computed over rolling windows, to cluster stocks into pairwise trading strategies or basket-based mean reversion portfolios. Recent clustering approaches improve market structure representation by moving beyond pairwise approaches to graph-based clustering that models equities as nodes in a network connected by edges weighted via dependence measures. Algorithms such as SPONGEsym exploit the global structure of correlation networks to form long-short baskets that provide natural hedging and diversification (Korniejczuk and Slepaczuk, 2024).

A fundamental limitation of these approaches is their reliance on backward-looking dependence measures. Rolling and exponentially weighted correlation estimates are inherently lagged because they average the past, therefore they continue reflecting previous regimes even after conditions shift. In practice, this produces inaccurate clusters formed on old information as actual correlations shift in days, whereas windowed estimators adjust after weeks. For instance, during the COVID-19 crisis equity correlations spiked sharply within days shifting the market structure and collapsing previously stable clusters into a small number of highly correlated groups, undermining the effectiveness fo clustering-based strategies (Packham \& Woebbeking, 2021). In such environments, basket construction based on lagged dependence structures exposes trading strategies to risk and persistent trading losses. Therefore, with increased market efficiency, greater competition, and structural shifts in asset dependence networks, there is growing motivation to move beyond these backward-looking dependence measures towards predicted dependence measures (Ozybayoglu et al., 2020). 

This paper therefore addresses the limitation of backward-looking dependence measures by introducing forward-looking correlation forecasts directly into the clustering stage of statistical arbitrage. Rather than predicting co-movement on historically formed baskets, we predict future correlations and construct clusters based on anticipated dependence structures. Forecasting correlations at short horizons is challenging as correlations are noisy, high-dimensional, and sensitive to regime changes driven by macroeconomic shocks, sector dynamics, and market-wide volatility. Capturing these dynamics requires models that can jointly represent temporal patterns at the asset level and relational structure across the market.

Recently, deep learning architectures have shown a remarkable ability to model complex, non-linear, and time-varying relationships in financial data (Zhang et al., 2020). Therefore, we propose a hybrid Transformer-Graph Attention Network architecture to forecast 10-day-ahead stock-stock correlations for S\&P 500 constituents. The model first encodes asset-specific temporal information using a Transformer, producing context-aware embeddings from recent return, technical, and macroeconomic features,. These embeddings are propagated through an edge-aware graph attention network that models cross-asset interactions and adapts to evolving network structure. To stabilize learning and anchor predictions to economically meaningful benchmarks, correlations are modeled in residual Fisher-z space relative to a rolling historical baseline.

The primary contribution of this study is to show that learned, forward-looking correlation forecasts can materially improve dependence estimation relative to rolling-window methods, particularly during periods of structural change. A second contribution is to demonstrate that incorporating these forecasts into graph-based clustering frameworks yields more adaptable baskets, mitigating the inherent lag of backward-looking approaches and improving basket trading returns and risk. Finally, we provide evidence that the attention mechanisms within the model offer insights into the importance of macroeconomic, sectoral, and return-based features across varying market regimes.

\section{Literature Review}

\subsection{Clustering Techniques for Statistical Arbitrage}
Identifying groups of equities with strong co-movement has a long history in statistical arbitrage (StatArb). Early approaches used economic sector classifications, such as Fama and French’s 48 groups based on industrial codes (Fama and French, 1997), which – although created for asset pricing researching – was used to group stocks for cross-sectorial arbitrage strategies. Pairs trading extended this concept through identifying stock pairs with historical co-movement (Gatev et al., 2006). When the spread of the two stocks increases, offsetting short/long positions (contrarian trading signals) are entered into with the expectation of mean reversion and thus, trading profits (Gatev et al, 2006). Basket-based StatArb generalizes pairs trading to small portfolios of co-moving stocks, typically improving diversification, reducing idiosyncratic risk, increasing capacity, and yielding more stable hedges than single-pair trades (Avellaneda \& Lee, 2008).

To identify pairs or groups of similar stocks a variety of methods have been commonly used over the last 25 years. Namely, correlation (Chen et al., 2019), Euclidean distance (Gatev et al., 2006), cointegration (Mikkelsen, 2011), Hurst Exponent (Bui and Ślepaczuk, 2022), or combinations of the metrics (Rad et al., 2014). While effective historically, profitability has tended to decay with increased competition, structural market changes, and reduced arbitrage opportunities (Hong \& Hwang, 2021). This has motivated the search for more robust clustering techniques. 

A major advance in StatArb clustering has been the adoption of graph-based clustering methods. Rather than treating each asset pair independently, these approaches capture the market-wide network structure by modeling equities as nodes linked by edges that represent dependence, such as correlation or cointegration. In highly interconnected financial markets, this global perspective of the rich market structure is key for basket trading as two stocks may exhibit weak direct correlation, but their shared network connections may place them in the same financial cluster. (Mantegna, 1999; Tumminello et al., 2005)

Building on early hierarchical clustering methods such as, Minimum Spanning Tree (Mantegna, 1999) and Planar Maximally Filtered Graphs (Tumminello et al., 2005), the Signed Positive Over Negative Generalized Eigenproblem (SPONGE) clustering algorithm represents residual stock returns as a signed, weighted graph thus allowing for anti-correlated stocks in long/short portfolios to provide natural hedging opportunities (Cartea et al., 2023). SPONGE then solves a generalized Eigenproblem to embed stocks before clustering stocks into k baskets through k-means++ clustering (Cartea et al., 2023). Cartea et al. (2023) tested many algorithms including Spectral Clustering (Ng et al., 2002) and Signed Laplacian Clustering (Kunegis et al., 2010), with SPONGE performing best specifically, the symmetric SPONGE version, SPONGEsym.

Cartea et al. (2023) demonstrated the potential of graph-based clustering for stat arbitrage by constructing long-short baskets every 3 days using SPONGEsym in all S\&P 500 equities, reporting annualized returns of 12.2\% and a Sharpe ratio of 1.1. However, under modest transaction costs the profitability of the strategy falls sharply. Korniejczuk and Slepaczuk (2024) extended this framework by rebalancing portfolios every 10 days and filtering stocks through an ensemble of standard machine learning classifiers to rank long-short position candidates. Rather than trading on all possible basket constituents each rebalance trades on only the top 10\% of positions thereby reducing the transaction costs that erode gains. Despite these advances, both studies are fundamentally backward-looking, as clusters are formed from a historical correlation market structure. During crisis periods, such as COVID-19 shock, these stale clusters result in trading losses while rolling windows adjust to the changed market structure. This limitation motivates the use of predicted, forward-looking correlation estimates in real-world basket trading to form baskets on anticipated rather than realized market structure. 

\subsection{Machine Learning for Forward-Looking Stock--Stock Correlation}

Modeling cross-asset correlations is crucial for risk management, portfolio optimization, and statistical arbitrage (Markowitz, 1952). Traditional approaches such as multivariate GARCH (Engle, 2008) and factor models (Fama \& French 1993), forecast covariance relationships but assume restrictive static or linear relationships which are frequently violated in practice. A popular extension is the Dynamic Conditional Correlation (DCC) model which allows correlations to vary over time (Engle, 2008). However, these models often struggle with structural breaks, nonlinear dependencies, and regime shifts, particularly at short-medium horizons (Longing \& Solnik, 2001; Dong \& Yoon, 2017). 

In contrast, advanced deep learning architectures instead learn complex, time-varying correlation structures directly from data (Zhang et al., 2020). Hybrid methods, such as the DCDNN framework (NI \& Xu, 2023), uses a DCC-GARCH to produce baseline forecasts then train a neural network to correct its residuals. Similarly, a DCC-GARCH with LSTM modules captures temporal dependencies in correlations, achieving better risk detection during periods of crisis such as the COVID-19 shock (Chunge et al., 2024). These hybrid methods improve predictive accuracy and highlight the power of deep learning models in correlation prediction tasks when correlation predictions is treated as a supervised learning task. However, these approaches do not combine both the temporal and cross-sectional structure simultaneously as a result, neither class of models fully captures the joint-evolution of time-varying market conditions and the dynamic relational structure linking assets.

A natural advance in deep learning for correlation prediction is toward an architecture capable of learning both the temporal effects and the relational structure of financial markets. Together, the joint architecture is able to generate forward-looking regime aware correlation forecasts that adapt to evolving market structure, changing macroeconomic conditions, and overcome the lag and rigidity of traditional models. This contrasts with classical regime-switching approaches - such as regime-switching VARs (Krolzig, 2002) and hidden Markov models with volatility clustering (Ryden et al., 1998) - which impose discrete state transitions. By allowing regime effects to be incorporated implicitly through features and adaptive weighting mechanisms, joint-temporal relational models adjust correlation estimates smoothly and rapidly as regimes emerge and fade.

\subsection{Graph Neural Networks for Correlation Modeling}
Graph Neural Networks (GNNs) provide a natural fit for financial markets because they model the market-wide network structure rather than isolated pairs. Stocks are nodes and edges represent current/lagged information, such as past or predicted correlations (Chen et al., 2022), common factor exposures (Fama \& French, 1993), sector membership or other meaningful economic connections. Each node carries features such as historical returns, volatility, technical indicators (for example stocks sectors), factor modeling, and macro variables. GNN training is supervised to minimize a loss between predicted and realized dependencies thus, the model learns to anticipate where relationships are likely to move rather than reflecting the past (Hamilton et al., 2017). This structure allows the model to capture both pairwise dependencies as well broader network effects.

The core mechanism in GNNs is message passing; at each layer, a node updates its latent representation – encoded data capturing a nodes most important relationships and patterns – through aggregating information from its neighbors, with aggregation weighted by the strength or type of connecting edge (Hamilton et al., 2017). Over multiple layers, information moves through the network such that the embedding of a stock reflects its own features, those of its neighbors, and neighbors-of-neighbors (Wang et al., 2022). This allows GNNs to capture indirect effects such as, shared macro shocks, inter stock dependencies and the overall network structure, that pairwise correlations and other models cannot capture (Wang et al., 2022). 

Basic GNN layers aggregate neighbor information with predefined, fixed weights limiting their ability to adapt as the importance of cross-asset relationships changes over time. Graph Attention Networks (GATs) addresses this by learning context-dependent, per-edge weights via attention mechanisms, allowing more informative neighbors to contribute more strongly to predictions (Velićković et al., 2018). Multi-head attention further enhances this flexibility by learning multiple weighting schemes in parallel, enabling each head to capture distinct structural patterns such as short-horizon co-movement and regime dependent shifts (Velićković et al., 2018). GNNs without attention often employ rigid thresholds to adjust edge weightings and remove uninformative edges to make the model regime adaptable (Wang et al., 2022). Through employing attention the model starts from informed starting points (historical correlations and other edge attributes such as, sector codes) and learns data-driven adjustments to produce edge/attention weights tailored to the current market regime, improving robustness to noise and regime changes (Velićković et al., 2018). This is particularly valuable in financial markets, where the relevance of specific cross-asset relationships varies across market conditions. and reduces the risk of the model becoming overly dependent on a single learned representation pattern. 

Empirically, in finance tasks GAT-based models outperform Graph Convolution Networks (GCN) and Long Short-Term Memory (LSTM) baselines and deliver higher Sharpe ratios (Xiang et al., 2023; Sawhney et al., 2021). Currently, explicit correlation prediction has mostly used non-GNN architecture, such as ARIMA-LSTM (Choi, 2018) which directly highlights a gap in direct correlation forecasting with GNNs. GNN-based models have been used to forecast multivariate realized volatility/covariance by using a Graph-Transformer hybrid for short-term realized volatility on U.S. equities (Chen \& Robert, 2021) and other graph-based forecasting methods (Zhang et al., 2025). To our knowledge, no prior work directly learns short-medium term day ahead equity correlations with a GNN. Most current papers that directly forecast correlations focus on daily/short-term horizons. 

\subsection{Transformers for Temporal Financial Modeling}
Transformers are well suited for finance because of their ability to capture the long-range temporal dependencies of stocks. Transformers use self-attention mechanisms allowing them to dynamically focus on the most relevant historical patterns for prediction, enabling shifts in market regimes to be captured.

Transformers have transformed the landscape for sequence modeling by replacing recurrent neural networks (RNNs) and convolutional neural networks (CNNs) with an architecture built around self-attention (Vaswani, et al., 2017). RNN-based approaches such as, Long Short-Term Memory Networks (LSTMs), process sequences sequentially, passing hidden states from one time step to another leading to vanishing or exploding gradients (Vallarino, 2024). CNN-based sequence models capture local temporal patterns through convolutional filters – small learnable weight matrices - that move step by step over the time series to detect short-term structures such as, volatility bursts (Bai et al., 2018). However, filters must be stacked to process large time steps with enough granuality but this increases computation and forces long-range information to pass through many intermediate steps, which can degrade signal quality and lead to fading gradients (Bai et al., 2018).

Transformers address these limitations through self-attention mechanisms which allow the model to process entire T-day sequences in parallel and to directly learn dependencies between any two points in time. The model can therefore focus on varying historical contexts without relying on fixed lag structures. Further, transformers use parallel processing making them computationally efficient for large-scale financial datasets. Transformers also use multi-head self-attention which allows the learning of multiple sets of attention weights in parallel. This allows the model to process different types of temporal patterns, just as in the GAT multi-head attention, enhancing the expressive power of the model. (Vaswani et al., 2017)

In financial modeling, transformers set the pace for a range of task such as, return forecasting (Zeng et al., 2022; Mozaffari et al., 2024) and volatility prediction (Frank., 2023). Transformers self-attention can reweigh any part of the history and adjust as conditions change, making them especially well suited to correlation forecasting, where informative horizons vary and market regimes shift. 

\subsection{Temporal-Heterogeneous GNN (THGNN)}
Recently, GNNs and Transformers have been combined in a single architecture as GNNs specialize in learning the spatial structure of market relationships, while Transformers specialize in modeling the temporal evolution of those relationships. Xiang et al. (2023) employ a hybrid Temporal-Heterogeneous GNN (THGNN) that combines a Transformer temporal encoder with a graph attention (GAT) relational model to predict whether the next days stock price will move or down. The transformers produces context-aware, asset-specific embeddings from the full T-day input window and the GAT then focuses on the relational structure between assets, weighting neighbors via learned attention. This two stage-process ensures that relational inference is informed by rich, context-aware temporal representations. Multi-head attention in both components allows the model to capture diverse interactions between assets such as, relationships during stable vs crisis periods or the relationships between positive vs negative correlations, improving stability and capturing of multiple dynamics. Empirically, Xiang et al. report superior performance to Transformer, GAT, GRU, LSTM, and several hybrids across predictive accuracy ($57.9\%$ accuracy) and risk-adjusted return metrics (such as Sharpe, Calmar, information ratio) on their equity dataset. Although the target is price direction rather than correlation and the time horizon is short, THGNN's joint modeling of time-varying network structure and temporal dependence directly motivates its use for forward-looking correlation estimation.

\subsection{Deep Learning Interpretability}

Deep learning interpretability is increasingly important in finance, where models inform high-stakes trading decisions and risk management. Models must provide both predictive accuracy as well as an understanding of what inputs drive model outputs and whether the model behaves sensibly across regimes. Complex deep learning models, such as Transformers and GATs, are often criticized as black-box models that do not provide direct interpretability like linear models (Shrikumar et al, 2017).

Model-agnostic explanation methods, such as Local Interpretable Model-agnostic Explanations (LIME) (Riberio et al., 2016) and SHapley Additive exPlanations (SHAP) (Lundberg \& Lee, 2017), treat the model as a black box and infer explanations by probing input-output behavior. Whereas, gradient-based attribution models examine the internal gradients of the model to estimate each input feature's influence on the output. Gradient $\times$ Input multiplies the gradient value by the input value to consider the overall contribution of an input to the output (Shrikumar et al., 2019). Simple Gradient $\times$ Input can provide time-step level attributions for sequence models, whereas path approaches such as Integrated Gradients and SmoothGrad yield more stable, global importance scores at the cost of less instantaneous interpretability. 

Xiang et al (2023) employ interpretability in their THGNN through evaluating the weights in the temporal graph attention layers and the heterogeneous attention of the GAT model. They extract per-edge attention coefficients during message passing, aggregate them by node properties (daily return volatility) to study how volatile stocks influence neighbors, and average attention over heads to form stable importance summaries. This method is intrinsic and provides a clear interpretable framework to understand the THGNN. However, attention does not always equal causal importance (Jain \& Wallace, 2019). High attention weights do not guarantee that perturbing or removing a feature or neighbor will materially affect the prediction, especially in deep, multi-layer ML architectures where interactions are non-liner (Jain \& Wallace, 2019). Attention maps and gradient methods - that consider the full contribution of an input (Gradient $\times$ Input) - provide an informative summary of the model's internal weighting rather than formal causal attributions (Jain \& Wallace, 2019). 

\section{Methodology}

\subsection{Problem Framing}
Traditional stat arbitrage clustering approaches, whether sector-based classifications, correlation/cointegration methods or more recent graph-based methods like SPONGEsym, are fundamentally backward-looking. SPONGEsym specifically – as in Cartea et al.’s (2023) approach -, focuses on historical correlation matrices build from residual returns, which only adapt after the market structure has already shifted. In crisis regimes, such as the COVID-19 shock, these correlations spike rapidly, collapsing clusters into a few dominant groups and generating trading losses. Even extensions like ML-filtered SPONGEsym – where trades are only entered if they are in the highest 10\% of predicted probability of being profitable – remain subject to this lag, as their predictive element is applied after the cluster formation step, not before. 

This creates a clear research gap: the application of modern machine learning architectures to predict future correlations, enabling clustering to be performed on forward-looking relationships rather than historical relationships. By conditioning trades on predicted rather than lagged correlation structures, such models adapt more rapidly to crisis regimes minimizing losses and capture additional gains in stable periods, as trades are based on the anticipated market state rather than one that has already evolved. 

We target 10-day-ahead correlations among S\&P 500 constituents, aligned with the 10-day rebalance period in the strategy put forth by Korniejczuk and Slepaczuk (2024). With a shorter rebalance, as in the original SPONGEsym implementation from Cartea et al. (2023), significant transaction costs are incurred thus, eroding profits.

Forecasting correlations at 10 days is challenging as correlations are noisy, high-dimensional, and lack a stable parametric form. Further, correlations are high-variance over this period, events such as, earnings announcements, macro releases, or sector news, can quickly alter correlations and even flip their sign, and many macro and supply-chain shocks occur and operate with a lag (Dupor, 2023; Carvalho et al., 2020). As a result, the prediction problem is governed by an evolving network rather than a static snapshot. Therefore, the proposed architecture combines a Transformer (temporal encoder) and GAT (relational encoder) in a Temporal-Heterogenous GNN (THGNN) to predict the future correlation of S\&P 500 stocks. 

Formerly the model will be tasked with predicting 10-day forward correlation from a date $t$. For each eligible pair $(i,j)$ the target correlation is
\begin{equation*}
\begin{gathered}
\rho_{(i,j,t+10)} = \mathrm{corr}\!\left( r_{(i,t+10)}, r_{(j,t+10)} \right) \\
\text{where } r_{(x,t+10)} \text{ is the 10-day cumulative return series for stock } x \text{ from time } t.
\end{gathered}
\end{equation*}

Further, correlations are bounded and heteroskedastic - the sampling noise depends on the level of $\rho$, so equal changes near 0 and near $\pm 1$ are not comparable. Therefore to stabilize variance and work on an approximately linear, unbounded scale, we apply the we apply the Fisher transform, $z=atanh(\rho)$, and train the model to predict the residual, $\Delta z$, relative to a rolling 30-day historical baseline. Z-space predictions are mapped back to correlations via $tanh$ for portfolio construction. This residual formulation anchors forecasts to a well-calibrated reference while allowing the model to correct for shifts in network structure that the backward-looking baseline cannot anticipate. Working in Fisher-z space makes residuals additive and gradients better-behaved, which stabilizes optimization and reduces the model collapsing to predict the mean - the model must at least learn deviations from a non-trivial baseline.

\subsection{Model Architecture}
\subsubsection{Data}

For each S\&P stock $i$ at date $t$ a feature vector $x_{(i,t)} \in \mathbb{R}^F$ that includes the following 37 features is constructed. 

\begin{itemize}
    \item Price and volume measures: closing price (PRC), trading volume (VOL) 
    \item Technical indicators: short- and medium-term momentum (5, 20, and 60 day), short-term reversal (5 day), relative strength index ($RSI_{14}$) and average true range ($ATR_{14}$)
    \item Firm characteristics: market capitalization, book-to-market ratio
    \item Factor exposures: rolling betas to the Fame-French three factors (mkt, smb, hml)
    \item Macroeconomic and risk factors: excess market return, size (smb), value (hml), risk-free-rate, momentum factor (umd), crude oil price (DCOILWTICO), 10-year Treasury yield (DGS10), trade-weighted dollar index (DTWEXBGS), VIX, and GARCH (1,1) implied volatility
    \item Return measures: daily excess return, raw return, and SPY return
    \item Sector and industry codes: gsector, gsubind
    \item Correlation and volatility context: rolling correlations with the market (10, 21, and 63-day windows), average 21-day correlation with sector and subindustry, realized volatility with the sector (20-day) and subindustry (20-day), 10-day realized market volatility, and cross-sectional return dispersion.
\end{itemize}

Equity data is obtained from CRSP and Computstat via WRDS, and macroeconomic variables from FRED. All derived features are computed using standard definitions and constructed out-of-sample to avoid look-ahead bias. To ensure data quality, stocks are inlcuded in a given day's graph only if sufficient recent history is available for feature construction. Specifically, as the sequence length of the transformer is 30 days, if a stock has 30 days of data within 33 trading days it is included and if not, it is excluded from graph construction. This balances data completeness with coverage of the S\&P 500 universe while avoiding overly restrictive filters. 

Feature groups provide complementary information across time scales: return variables capture fast-moving dependence changes, correlation and volatility measures encode medium-horizon co-movement, and macroeconomic variables reflect slower-moving regime conditions that systemically reshape cross-asset relationships. Sector identifiers provide a structural baseline that can be dynamically reweighted by the attention mechanism as regimes evolve.

To ensure that the model is not dominated by large high-variance variables all features are normalized using a rolling 60-day z-score. A time-based 80-20 train-validation split is employed, with training data from 2006-2018 and evaluation on post-2019 data. To avoid leakage from rolling statistics, the out-of-sample period begins 60 trading days after the start of 2019. 

\subsubsection{Temporal Encoder (Transformer)}

For every stock $i$ at date $t$, we form a sequence $X_{(i,t)} \in \mathbb{R}^{L \times F}$ from the previous $L=30$ trading days of $F=37$ features. Each feature sequence is mapped to $d_{\mathrm{model}}=128$ via a learned linear projection, with sinusoidal positional encodings adding to preserve temporal order. The resulting $30 \times 128$ sequence is then passed through four pre-norm Transformer encoder layers with 8 attention heads and dropout set to 0.2 to avoid the risk of overfitting. Therefore each head has a key/query width $d_k=16$, $d_k = d_{\mathrm{model}} / n_{\mathrm{heads}} = 16$. 

Let $H_{(i,t)} \in \mathbb{R}^{L \times d_{\mathrm{model}}}$ denote the final encoder output. We then flatten the matrix $H_{(i,t)}$ to $\mathrm{vec}(H_{(i,t)}) \in \mathbb{R}^{L \cdot d_{\mathrm{model}}} = \mathbb{R}^{3840}$ and apply a LayerNorm-MLP to obtain a 512-dimensional node embedding $h_i^{(0)} \in \mathbb{R}^{512}$, which becomes the initial node features for the graph module (GAT). 

A model width $d_{model}=128$ is chosen as this provides sufficient capacity to encode the 37 heterogeneous features and their complex interactions, while remaining efficient and stable on daily data. Eight heads are utilized as this allows the model to specialize in different horizons or volatility regimes without fragmenting capacity cross too many subspaces. The resulting $d_k=16$ also keeps the scaled dot-products well formed. Flattening the $30 \times 128$ embeddings to a singular 512-d embedding compresses the rich temporal representations into a dimensional space that is expressive enough to preserve regime and pattern information. It minimizes overfitting and ensures that the GAT's memory and compute is manageable. These three architecture choices are also in line with the impressive performance achieved by the THGNN employed by Xiang et al. (2022) to determine if a stock price will increase or decrease. 

The original THGNN employed by Xiang et al. (2022) uses a single self-attention layer over a 20-day window. We extend both the temporal horizon and depth. First, we set $L=30$ because the model can capture a full month of short-term shocks, weekly effects, and month-end flows (often aligned with macro indicator releases) without diluting signal with stale history or incurring excessive compute. Longer windows that span multiple years theoretically allow the model to see similar market regimes, but they substantially increase training cost; we leave their evaluation to future work. Secondly, we use a 4-layer pre-norm Transformer to compound non-linear temporal patterns across the 30-day window. 1-2 layers risks underfitting this horizon, while deeper stacks increase compute and overfitting risk with likely minimal gains. This choice is validated by the strong out-of-sample performance and provides enough capacity to separate fast shocks from slower drifts and to model complex interactions between returns, volatility, and macro state, yet shallow enough to generalize.
 
\subsubsection{Graph Building}

At each $t$ an undirected, signed, weighted graph $G_t = (V, E_t, W_t)$ is built. Let $\rho^{base}_{(i,j,t)}$ be the previous 30-day rolling window correlation for stocks $i$ and $j$ at time $t$.

\begin{itemize}
    \item \textbf{Edge sampling:} for every stock the top 50, bottom 50 correlations, and 75 randomly sampled mid-strength partners - sampled from $[0.2,0.8]$ correlation percentiles - by $\rho^{\mathrm{base}}$ are included as weighted edges.
    \item \textbf{Weights:} $w_{(ij,t)} = \rho^{\mathrm{base}}_{(ij,t)} \in [-1,1]$.
    \item \textbf{Edge attributes:} $a_{(ij,t)}$: baseline correlation $\rho^{\mathrm{base}}_{(ij,t)}$, $|\rho^{\mathrm{base}}_{(ij,t)}|$, and sign indicator (0 if $\rho^{\mathrm{base}}_{(ij,t)} > 0$ and 1 otherwise). We also include binary edge flags for the same sector (gsector) and same sub-industry (gsubind) (1 if same, 0 otherwise).
    \item \textbf{Relation class:} a discrete label for low/neutral/high correlations used as an edge-type embedding in the GAT attention mechanism. The bottom 3rd of $\rho^{\mathrm{base}}_{(ij,t)}$ are 0, the middle 3rd of $\rho^{\mathrm{base}}_{(ij,t)}$ are 1, and the top 3rd of $\rho^{\mathrm{base}}_{(ij,t)}$ are 2.
    \item \textbf{Features:} all stocks learned 512-dimensional node embeddings $h_i^{(0)}$.
\end{itemize}

This yields a signed, feature-rich stock network that combines rich node embeddings with informative edge attributes, capturing both positive and negative co-movements. The sampling scheme balances exposure to strong positive, strong negative, and ambiguous relationships, without unnecessarily increasing the compute time of training.

\subsubsection{Relational Encoder (GAT)}
Following temporal encoding, a graph attention network (GAT) models cross-asset interactions. At each date $t$, node embeddings produced by the Transformer serve as initial graph features. The relational encoder updates these representations through edge-aware, multi-head attention, allowing each stock to selectively aggregate information from economically relevant neighbors.

We stack $L_g = 3$ graph attention layers, each with $H =4$ attention heads. At each layers $l$, node states $h_i^{(l)} \in \mathbb{R}^{512}$ are projected independently per head using learned linear maps, yielding head-specific representations that learn different relational patterns. 

For edge $(i,j)$, head $h$, and layer $l$, we form an edge-conditioned gate
\[
m_{(i,j)}^{(l,h)} = E^{(h)}_{\text{type}}(\tau_{ij}) + W_f^{h} f_{ij}+W_a^{(h)} a_{(ij,t)} + W_s^{(h)}e_{ij}^{(l)},
\]
where $\tau_{ij} \in \{\text{neg}, \text{mid}, \text{pos}\}=\{0,1,2\}$ denotes the baseline correlation regime (negative, neutral, positive), $f_{ij}$ encodes sector and sub-industry similarity via is a 2-dim binary vector, $a_{(ij,t)}$ are the edge attributes, $e_{ij}^{(l)}$ is a persistent edge state that propagates relational context across layers.

For each head, attention coefficients are computed from the concatenated projected node features and edge gate via LeakyReLU, normalized over the neighborhood of node $i$ via softmax, and used to aggregate neighbor representations. Outputs from all heads are concatenated, linearly projected, and combined with residual connections and layer normalization to produce updated node embeddings $h_i^{(l+1)}$. In parallel, edge states are updated via a residual MLP conditioned on the updated endpoint embeddings, allowing higher-order relational information to accumulate across layers.

After the GAT layer, node and edge representations are combined to form pairwise edge embeddings,
\[
u_{ij}^{\text{edge}} = [h_i^{(L_g)} \, || \, h_j^{(L_g)} \, || \, e_{ij}^{(L_g)}].
\]

We route $u_{ij}^{\text{edge}}$ by the edge types, $\tau_{ij}=\{neg, mid, pos\}=\{0,1,2\}$, to one of three small MLP expert heads which allows the model to consider the different dynamics of correlation regimes. This is very important as correlation prediction is noisy and negative correlations behave very differently than positive correlations; therefore, the relationships underpinning different correlations values are different. Each head outputs a scalar residual in Fisher $z$-space, $\Delta \hat{z}_{ij}$, that is added to to the historical baseline and mapped backed to correlation space, 
\[
\hat{\rho}_{ij} \;=\; \tanh\!\big(z^{\text{base}}_{ij} \;+\; \Delta \hat{z}_{ij}\big)
\]

In line with Xiang et al. (2022) a 512 dimension embedding is used as this is wide enough to encode heterogeneous drivers (returns, macro, structure) and small enough to keep compute time and memory manageable. We also use four self-attention heads per layer. More heads are used in the temporal encoder as with a sequence of 30 days we are able to efficiently disentangle overlapping phenomena - very recent shocks, weekly patterns, slower drift - at a manageable computational cost of $O(L^2)$. Contrastingly, the GAT operates on dense edge graphs where each node scores many neighbors and attention is conditioned on node features, edge attributes, and a persistent state. In this setting,  edge-conditioned. Therefore, we prioritize per-head capacity over head count as 4 wider heads yields a more expressive, fewer heads with higher per-head dimensionality provides a favorable between expressive power and stability of attention weights, while keeping runtime manageable. Additionally, given the edge wise routing to neg/mid/pos heads specialization already exists therefore, splitting into 8 graph heads would likely provide little benefit.

Similar to the transformer, we increase the depth of the model from one attention layer used by Xiang et al. (2022) to three. Deep stacks would overfit daily graphs and increase compute, while a single layer can underfit cross-asset interactions. Three layers provides a practical middle ground that allows sufficient capacity for information to flow between layers and keeps compute time manageable. 

Overall, this THGNN architecture allows the model to learn the complex interactions between variables and model the relationships that govern negative, neutral, and positive correlations, whilst balancing compute time, stability, and memory. 

\subsubsection{Loss Function and Fisher z-space}
\label{subsubsec:loss-fisher}

The model should be trained to minimize the error between the predicted 10-day return correlation and the actual observed future 10-day return correlation for a stock pairing $i,j$. Therefore, we employ a Smooth-L1 (Huber) loss function that penalizes the difference between the predicted and realized Fisher values:
\[
L_{\text{edge}} = \mathrm{Huber}(\hat{z}_{(ij,t+10)}, z_{(i,j,t+10)}).
\]
Huber is less sensitive than MSE to large residuals as it behaves quadratically for small residuals and linearly for large ones, giving stable gradients near zero while being robust to regime shocks that create large errors.

During initial training using only Huber edge loss, $L_{\text{edge}}$, the model experiences mode collapse where predictions from each edge type bunched in narrow bands. This is common for deep models that are trained on noisy targets, as the easiest way to reduce pointwise loss is to predict values near the mean of each head, sacrificing spread and tails. That behavior is fatal for our use-case, because clustering and portfolio construction depend on the shape of the correlation distribution, not just the average accuracy per pair. If everything sits neat the mean of each head, clusters are blurred therefore, the clusters formed are uninformative thus the model has limited use case and does not model the complex interactions which govern correlations. 

Therefore, we frame learning as residual prediction around a strong up-to-date baseline. Let $\rho_{(i j,t)}^{base}$ denote the correlation between stocks $i$ and $j$ computed on a rolling 30-day window ending at time $t$. The model is trained to predict the change, $\Delta \rho_{(i j,t )}$. We predict this residual in Fisher $z$-space. 
\[z_{(i j,t)}^{base}=\mathrm{atanh}(\rho_{(i j,t)}^{base})\quad \text{and} \quad \hat{z}_{(ij,t\to t+10 )}=\mathrm{atanh}(\hat{p}_{(i j,t\to t+10 )}).\] 
The model outputs $\Delta\hat{z}_{(ij,t)}$ such that
\[\hat{z}_{(ij,t\to t+10 )}=z_{(i j,t)}^{base}+\Delta\hat{z}_{(ij,t)} \quad \text{and} \quad \hat{p}_{(ij,t\to t+10 )}=\tanh(\hat{z}_{(ij,t\to t+10 )}).\] 
This ensures that the model at a minimum, learn deviations from a non-trivial baseline thus, negating the model predicting the mean. Further, this also makes gradients additive and linear, making gradients well-behaved (no vanishing or exploding gradients) and increasing model stability.

Moreover, a histogram-matching loss term which forces the network to reproduce the distribution of correlations across the market rather than bunching predictions into narrow bands was added. Intuitively, Huber teaches the model to make edge prediction as close to its true value, while the histogram term teaches the model to make the whole market look realistic. This ensures that the relative ordering of stock correlations and their separation is preserved.

To allow the model to update weights based on both distribution matching and pointwise edge loss, both loss functions must contribute gradients. Therefore, hard binning histogram binning cannot be used as this contributes no gradients. As such, Gaussian soft binning will be used as this is differentiable. Gaussian soft binning assigns each sample to nearby bin centers with Gaussian weights. Each correlation is given a weight for each bin with a higher weighting given to bins it is closer to. (Karandikar et al., 2021)

Let $x_1,\dots,x_N$ be a batch of predicted correlations and $c_1,\dots,c_B$ be bin centers spanning the batch target range. Each sample contributes to nearby bins and is normalized via
\[
w_{(n,b)} = \exp\!\left(-\frac{(x_n - c_b)^2}{2\sigma^2}\right), 
\qquad
h_b = \frac{1}{N} \sum_n \frac{w_{(n,b)}}{\sum_{b'} w_{(n,b')}} ,
\]
where $b^{'}$ is the bins for the sample $n$.

The histogram loss function is therefore:
\[
L_{\text{hist}} = \| h_{\text{pred}} - h_{\text{true}} \|_2^2 = \sum_{b=1}^B (h_{\text{pred}}[b] - h_{\text{true}}[b])^2.
\]

To ensure that each of the three edge types properly matches the distributions we compute the histogram losses for each separate edge type using 6 bins and a global histogram loss of all correlations using 15 bins. Each of the four histogram losses are given equal weights and the overall histogram loss is multiplied by a constant factor of $s=7$ to ensure that it is comparable in scale to the edge loss term, so that the model will optimize both loss functions.

The total loss function gives equal weights to both the histogram and edge losses.

\subsubsection{Overview}
\begin{figure}[H]
\begin{center}
\begin{tikzpicture}[
  font=\scriptsize,
  >=Latex,
  node distance=4mm and 12mm,
  block/.style={draw, rounded corners, align=center, inner sep=2pt, thick, fill=white},
  group/.style={draw, rounded corners, dashed, inner sep=6pt, thick},
  decision/.style={draw, diamond, aspect=2, inner sep=1pt, thick, align=center},
  line/.style={-Latex, thick, shorten >=1pt, shorten <=1pt},
  lbl/.style={fill=white, inner sep=1pt, text depth=0pt, text height=1.5ex},
  title/.style={font=\bfseries}
]

\node[block, minimum width=40mm] (infeat) {Per-stock input\\ 30x37 (60d rolling z-score)};
\node[block, below=of infeat, minimum width=40mm] (proj) {Linear projection\\ $d_{\mathrm{model}}=128$};
\node[block, below=of proj, minimum width=40mm] (posenc) {Add positional encodings};
\node[block, below=of posenc, minimum width=40mm, text width=45mm] (attn) {4x Pre-Norm Self-Attention\\ (8 heads)};
\node[block, below=of attn, minimum width=40mm] (seqout) {Sequence out\\ $30 \times 128$};
\node[block, below=of seqout, minimum width=40mm] (flat) {Flatten};
\node[block, below=of flat, minimum width=40mm] (nodeemb) {MLP $\rightarrow$ 512-d node embedding};

\node[group, fit=(infeat)(proj)(posenc)(attn)(seqout)(flat)(nodeemb),
      label={[title]above:Per-Stock Transformer Encoder}] (encgrp) {};

\draw[line] (infeat) -- (proj);
\draw[line] (proj) -- (posenc);
\draw[line] (posenc) -- (attn);
\draw[line] (attn) -- (seqout);
\draw[line] (seqout) -- (flat);
\draw[line] (flat) -- (nodeemb);

\node[block, right=30mm of attn, minimum width=36mm] (basecorr) {Baseline correlations\\ (30d)};
\node[block, below=of basecorr, text width=46mm] (edgepick) {Edge selection (per node):\\ 75 top corr + 75 bottom corr + 50 random mid};
\node[block, below=of edgepick, text width=48mm] (edgeattr) {Edge attributes:\\ corr, $|$corr$|$, sign, same-sector (gsubind),\\ relation type: neg/mid/pos $\to$ 0/1/2};

\node[group, fit=(basecorr)(edgepick)(edgeattr),
      label={[title]above:Graph Construction (Day $t$)}] (graphgrp) {};

\draw[line] (basecorr) -- (edgepick);
\draw[line] (edgepick) -- (edgeattr);

\node[block, below=18mm of nodeemb, minimum width=50mm, text width=54mm] (gat)
  {Edge-aware Multi-Head GAT\\ (3 layers, 4 heads)};

\node[group, fit=(gat),
      label={[title]above:Message Passing}] (gatgrp) {};

\draw[line] (nodeemb) -- node[lbl, above, pos=0.55]{node emb (512d)} (gat);
\path (edgeattr.west) -- ($(edgeattr.west)+(-14mm,0)$) coordinate (edgebend);
\draw[line] (edgeattr.west) to[out=180,in=90] (edgebend)
           to[out=-90,in=20] node[lbl, above, pos=0.65]{edge feats} (gat.north east);

\node[block, below=10mm of gat, minimum width=40mm] (edgeh) {Edge representations $h_{ij}$};
\draw[line] (gat) -- (edgeh);

\node[decision, below=10mm of edgeh] (route) {Relation type};
\draw[line] (edgeh) -- (route);

\node[block, below left=10mm and 28mm of route, text width=34mm] (expneg) {Expert MLP\_neg\\ $\to \Delta z_{ij}$};
\node[block, below=12mm of route, text width=34mm] (expmid) {Expert MLP\_mid\\ $\to \Delta z_{ij}$};
\node[block, below right=10mm and 28mm of route, text width=34mm] (exppos) {Expert MLP\_pos\\ $\to \Delta z_{ij}$};

\node[group, fit=(route)(expneg)(expmid)(exppos),
      label={[title]above:Relation-Routed Experts}] (expgrp) {};

\draw[line] (route) -- node[lbl, above, pos=0.35]{0: neg} (expneg);
\draw[line] (route) -- node[lbl, right, pos=0.55]{1: mid} (expmid);
\draw[line] (route) -- node[lbl, above, pos=0.35]{2: pos} (exppos);

\node[block, below=18mm of expmid, text width=46mm] (prez)
  {$z^{\mathrm{pred}}_{ij} = z^{\mathrm{base}}_{ij} + \Delta z_{ij}$};

\draw[line] (expneg.south) to[out=-90,in=180] (prez.west);
\draw[line] (expmid.south) -- (prez.north);
\draw[line] (exppos.south) to[out=-90,in=0] (prez.east);

\node[block, right=28mm of prez, text width=36mm] (basez) {Baseline Fisher-$z$};
\coordinate (basez-mid) at ($(basez.north)+(0,15mm)$);
\draw[line] (basecorr.east) -| ($(basecorr.east)+(8mm,0)$) |- (basez-mid) -- (basez.north);
\draw[line] (basez.west) to[out=180,in=0] node[lbl, above, pos=0.5]{add} (prez.east);

\node[block, below=10mm of prez, text width=40mm] (tanh) {$\rho_{ij} = \tanh\!\big(z^{\mathrm{pred}}_{ij}\big)$};
\node[block, below=8mm of tanh, text width=44mm] (mat) {Forward-looking correlation matrix};
\node[block, below=8mm of mat, text width=48mm] (spongesym) {SPONGEsym clustering \& trading};

\draw[line] (prez) -- (tanh);
\draw[line] (tanh) -- (mat);
\draw[line] (mat) -- (spongesym);

\end{tikzpicture}
\end{center}
\caption{Proposed Transformer–GAT (THGNN) architecture for 10-day forward stock-stock correlation prediction. The model couples a Transformer encoder, which summarizes each stock's recent history into a latent representation, with a relation aware GAT that propagates information across the S\&P 500 using correlation and sector-based edges.}
\label{fig:thgnn-architecture}

\end{figure}
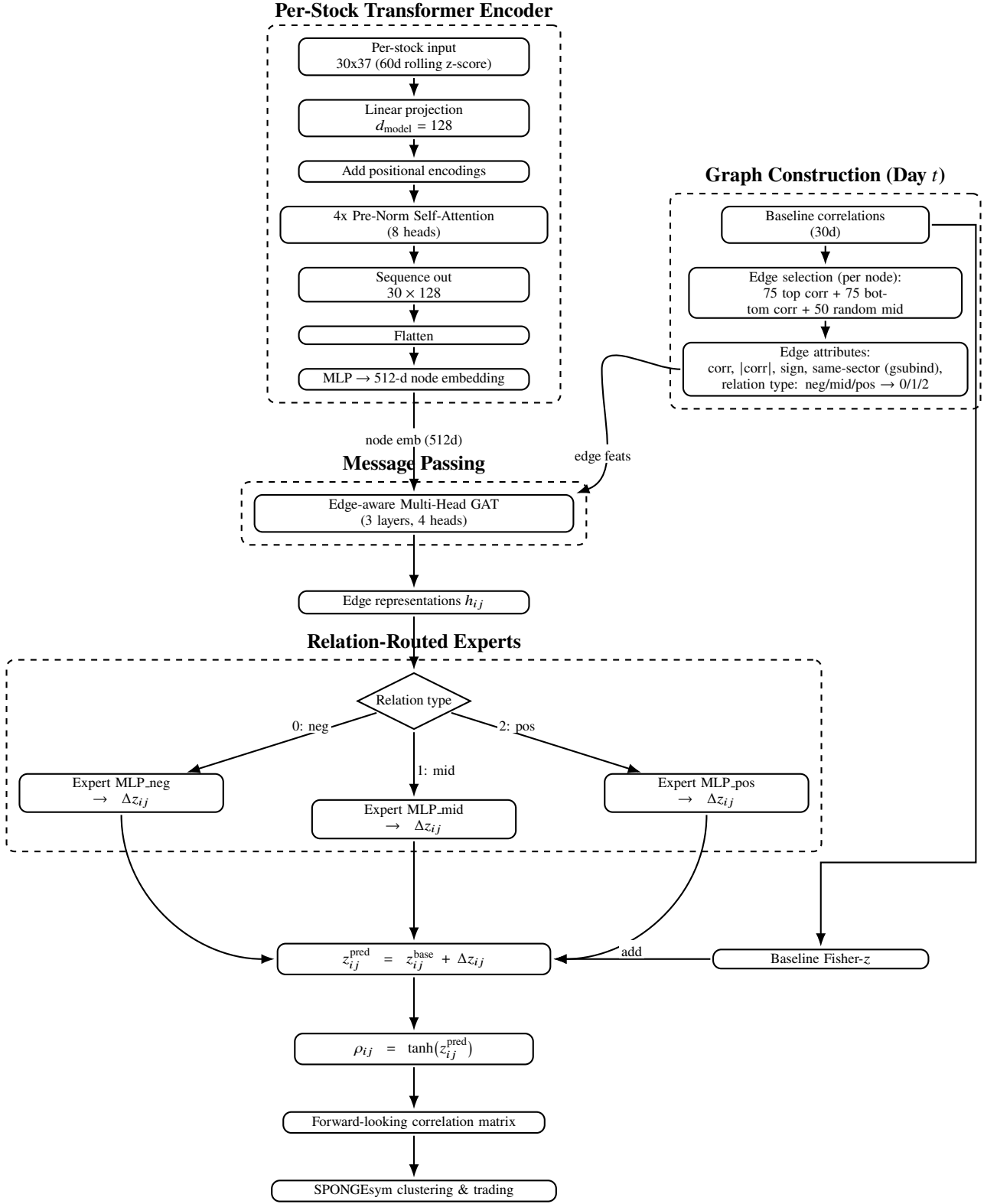

The proposed architecture is summarized in Figure \ref{fig:thgnn-architecture}. It integrates a temporal Transformer encoder with a relation-aware GAT to capture both sequential and cross-sector dependencies among S\&P 500 stocks. The Transformer processes each stock's historical feature sequence to produce latent node embeddings, while the graph module models interactions through edge attributes such as pairwise correlation. These embeddings are routed to specialized expert heads based on the historical correlation values, producing forward looking Fisher-$z$ residuals. The predicted residuals are added to the 30-day historical baseline, mapped back to correlation space via the hyperbolic tangent, and supplied to the SPONGEsym clustering and trading framework. 

The model is trained using AdamW with a cosine learning-rate schedule and gradient clipping to ensure stability in the presence of noisy, non-stationary financial targets. Due to the size of daily stock graphs, training is performed with small batches and gradient accumulation to achieve an effective batch size of 18. Standard pre-norm Transformer and GAT architectures are used throughout, and training is conducted over 75 epochs. See Appendix \ref{Training} for additional training details.

\subsection{Trading}

Trading decisions in this study follow the framework outlined by Korniejczuk and Slepaczuk (2024). Despite the inability to reproduce the the results obtained by Korniejczuk and Slepaczuk (2024) (see section \ref{Replication}), their framework remains well motivated. SPONGEsym provides a robust mechanism for forming long/short baskets and ML classifier filtering focuses on the highest-confidence signals, helping to manage turnover and transaction costs. Therefore, we retain their trading engine but use forward-looking predicted correlation matrices from the THGNN, instead of backward-looking rolling-window correlations. This ensures that portfolio allocation is informed by predicted rather than realized dependency structures.  

The SPONGE algorithm creates an adjacency matrix, $A$, from the correlation matrix and decomposes it into positive ($A^+$) and negative ($A^-$) components, from which positive and negative Laplacian matrices are constructed. The Laplacians are then normalized to $L_{\mathrm{sym}}^+$ and $L_{\mathrm{sym}}^-$,
\[
L_{\mathrm{sym}}^+ = (D^+)^{-1/2} L^+ (D^+)^{-1/2}, 
\quad 
L_{\mathrm{sym}}^- = (D^-)^{-1/2} L^- (D^-)^{-1/2}.
\]

SPONGE then solves the generalized eigenvalue problem by finding the $k$ smallest eigenvalues that satisfy
\[
(L_{\mathrm{sym}}^+ + \tau^- I) v = \lambda (L_{\mathrm{sym}}^- + \tau^+ I) v,
\]
where $\tau^-$ and $\tau^+$ are regularization parameters and $I$ is the identity matrix. This eigenvalue decomposition maps stocks into a $k$-dimensional Euclidean space, where traditional $k$-means++ clustering is applied to form clusters. The explicit modeling of the positive and negative dependencies allows SPONGE to create richer, more interpretable clusters than other methods such as Spectral Clustering (ng et al., 2002) and Signed Laplacian Clustering (Kunegis et al., 2010).

For a given trading between 2019 and 2024 the THGNN produces a full $N \times N$ predicted correlation matrix across the S\&P500 constituents. SPONGE-sym clustering (Cartea et al., 2023) is then applied to generate baskets of stocks with the optimal number of clusters, $k$, determined by the 90\% explained variance eigenvalue criterion on the symmetrized similarity matrix to ensure the number of clusters adapt with changing market regimes - particularly important during crisis periods - (Cartea et al., 2023). Let $\lambda_1 \geq \lambda_2\geq\dots\geq\lambda_N$ be the eigen values. We pick the smallest k such that
\[\frac{\sum_{i=1}^k \lambda_i}{\sum_{i=1}^N \lambda_i}\geq 0.90\],
and use the top-k eigenvectors as the spectral embedding fed to k-means to obtain the final clusters.

Within each of the $k$ stock clusters equal contrarian weights are generated. Long positions are taken if a stock underperforms the cluster mean cumulative return over the last 5 days and short positions are taken if a stock outperforms. To control transaction costs stock positions are rebalanced every 10 days.

A key innovation presented by Korniejczuk and Slepaczuk is the use of signal filtering. For each potential long or short candidate, five ML classifiers - Histogram Gradient Boosting (HGB), AdaBoost, Multi-layer Perceptron (MLP), Logistic Regression (LogReg), and Stochastic Gradient Descent (SGD) - are trained on binary indicators of profitability derived from back testing the baseline clustering strategy. Two alternative profitability definitions are used when labeling the training data:
\begin{itemize}
    \item \textbf{Threshold-based profitability:} A signal is considered profitable if its cumulative return over the holding period exceeds a fixed-take profit threshold of 4\% at any time during the 10-day holding period (as 4\% yields a balanced split between profitable and unprofitable trades). During the trading period any trade that reaches 4\% profit at any point will be closed out as this is considered a signal of completed mean reversion.
    \item \textbf{Transaction-cost-adjusted profitability:} A signal is considered profitable if, even without reaching the threshold, its return exceeds the transaction cost incurred at the next rebalance, thereby avoiding a loss.
\end{itemize}

Each signal in the training set is thus labeled profitable or unprofitable if either of the above conditions are met. The models are trained via a grid search under the feature set proposed by Korniejczuk and Slepaczuk (2024). See Appendix \ref{MLFiltering} for details. 

The outputs of the 5 models are aggregated through a weighted soft-voting ensemble with the model weightings determined by model performance. Model performance is measured using Brier Score Loss, a scoring rule that measures the mean squared error between predicted probabilities and actual binary outcomes. Unlike accuracy, which only reflects class assignment, the Brier Score evaluates the quality of probability estimates therefore, making it ideal for trading applications. 

To ensure both cost efficiency and diversification, only the top 10\% of signals by predicted probability of profitability are traded, yielding an average portfolio size of 45 stocks. A portfolio with 40-50 stocks provides effective diversification, with diminishing diversification benefits beyond this point (Raju, 2021). 

\subsection{Model Performance Evaluation}

The model’s performance will be evaluated both on its improvement to predicting future 10-day correlations and its trading performance.

To evaluate the model’s predictive performance the MAE, RMSE, Bias (mean signed error), and both the Pearson and Spearman correlations between predicted and actual future values of the model throughout the out of sample period from 2019-2024 will be compared to a baseline MAE through this period, where previous 20-day rolling correlations are used as future 10-day correlations. This will allow us to see if the model has greater predictive power then simply taking that future correlation will be past correlation. 

The trading performance of the model will be assessed through comparing the performance to the S\&P500 buy-and-hold benchmark. An equity curve will be plotted to understand the model's performance during varying macroeconomic regimes and the model will be evaluated based on the following metrics: Annualized Return (ARC), Annualized Standard Deviation (ASD), Sharpe Ratio, Sortino Ratio, Maximum Drawdown (MDD), Maximum Loss Duration (MLD), Calmar Ratio (CR), and Information Ratio (IR).

Furthermore, we will test if the Sharpe ratio for the strategy significantly outperforms the Sharpe ratio of the S\&P 500. Specifically, we aim to test,
\[\Delta S = S_{strategy} - S_{S\&P 500}\]
\[H_0: \Delta S = 0 \text{ vs } H1:\Delta S \neq 0\]
Given that equity returns display autocorrelation, conditional heteroskedasticity, and heavy tails, many traditional significance assumptions tests are violated (Tseng \& Li, 2011). Therefore, if the Ljung-Box test (autocorrelation) (Ljung \& Box, 1978), ARCH LM test (conditional heteroskedasticity) (Engle, 1982), or the Shapiro-Wilk test (non-normality) (Shaprio \& Wilk, 1965) are violated then parametric Sharpe tests are not appropriate. See Appendix \ref{Significance} for more details.

In such case, to test if the Sharpe ratios are significantly different bootstrapping can be used as this requires no assumptions about parametric returns. We employ the Stationary Bootstrap (SB) to preserve serial dependence while avoiding sensitivity to a fixed block length (Politis \& Romano, 1994). To improve coverage under heteroskedasticity and heavy tails, a studentized CI is also computed (Politis \& White, 2004). See Appendix \ref{Significance} for more details.

\subsection{Interpretability of the Transformer}
To evaluate feature-level interpretability within the Transformer, we cannot directly look at attention weights. In building embeddings the transformer creates a latent embedding of 128 from a feature set of 37 whereby, each latent coordinate is a mixture of the complex non-linear interactions between the 37 features determined by the learned weights over four stacked transformer layers. Therefore, we cannot directly interpret the attention weights as they operate in latent space, not on raw features. As such, to interpret the importance of each feature in building the latent embeddings we adopt Gradient $\times$ Input saliency scores, which multiply the gradient of the output with each input to provide a complete importance value for each feature (Shrikumar et al., 2019).

For each trading day in the out-of-sample period (2019-03-26 to 2014-12-05), we pass the sector-stock graph through the full Transformer-GNN model and define a target as the mean of the predicted edge correlations. We then backpropagate gradients from this target to the model's input features. The derivatives with respect to each input dimension are then multiplied by the corresponding input values which yields Gradient $\times$ Input scores. This captures the sensitivity of the model's prediction to small changes in each feature thereby representing how much the model's output would be expected to change in response to changes of a given predictor. To move from node-level scores to sector-level interpretability, we average Gradient $\times$ Input across all stocks within a sector. This produces, for each day, a vector of feature importance within a sector. These scores are normalized to sum to 1, recorded across time, and represented as heatmaps and time series, allowing persistent drivers of the Transformer's embeddings and regime-dependent shifts in feature reliance to be evaluated.

\subsection{Intepretability of the Graph Attention Network}
For the GAT component of our architecture, interpretability arises directly from the attention mechanism. In each message-pass the model builds the representation of stock $i$ as a weighted aggregation of its neighbors' embeddings, where the attention coefficient $a_{ij}$ controls the contribution of neighbor $j$ to the update of $i$. Unlike the Transformer case, these coefficients act directly on the node embeddings in the final prediction, so they can meaningfully be interpreted as relative influence weights across stocks. 

Since the network employs multi-head attention and multiple forward passes, we record the attention weights for each edge throughout the out-of-sample period and average across heads to obtain a single importance score per edge. This produces a daily time series of stock-to-stock attention weights, which are aggregated to the sector level by summing over all edges from firms in sector $X$ to those in sector $Y$. These summaries allow us to examine how the model reallocates attention across sectors during different market regimes, offering insight into how systemic shocks propagate through the equity network. While attention weights do not imply causality, they provide a transparent window into how the model prioritizes relational information under stress versus stable conditions.

\section{Results}

\subsection{Correlation Prediction Performance}

Table \ref{tab:metrics-side-by-side} reports measures for the correlation prediction accuracy when using the model's predicted future 10-day correlations on the out-of-sample period versus taking the historical 20-day correlations as future values. The predictions are evaluated for $269,121,442$ edges therefore, given the scale of the evaluation sample and the magnitude of changes, results can be considered economically meaningful.

\begin{table}[H]
\caption{Comparison of edge-wise correlation prediction accuracy for future 10-day stock-to-stock correlations between the baseline historical rolling window estimates and the predicted THGNN correlations.}
\label{tab:metrics-side-by-side}
\centering
\begin{tabular}{lrr}
\toprule
\textbf{Metric} & \textbf{Persistence Baseline} & \textbf{Model Predictions} \\
\midrule
Edges evaluated & 269{,}121{,}442 & 269{,}121{,}442 \\
MAE & 0.3071 & 0.2302 \\
RMSE & 0.3852 & 0.2940 \\
Bias (mean signed error) & $-0.0049$ & $-0.0012$ \\
Pearson $r$ & 0.310 & 0.778 \\
Spearman $\rho$ & 0.314 & 0.795 \\
\bottomrule
\end{tabular}
\end{table}

The model generates a significant decline of $0.0769$ in MAE as well as a decline of $0.0912$ in RMSE. These reductions indicate the model has substantially smaller deviations between predicted and realized correlations. 

Measures of correlation improve even more impressively. The Pearson correlation improves from $0.310$ to $0.778$, increasing the fraction of variance explained in future 10-day correlations from 9.6\% to 60.5\%. Further, the Spearman rank correlation increases from $0.314$ to $0.795$, indicating much tighter alignment in the rank ordering of predicted values versus realized correlations.

All together, the results indicate that the model greatly outperforms the naive historical correlation method, delivering accurate predictions that also agree with the realized structure of correlations. This combination is precisely what is required for downstream tasks such as, clustering and portfolio construction where both magnitudes and ranking of correlations affect clustering and weightings.

\begin{figure}[H] 
    \centering
    \includegraphics[width=1.0\textwidth]{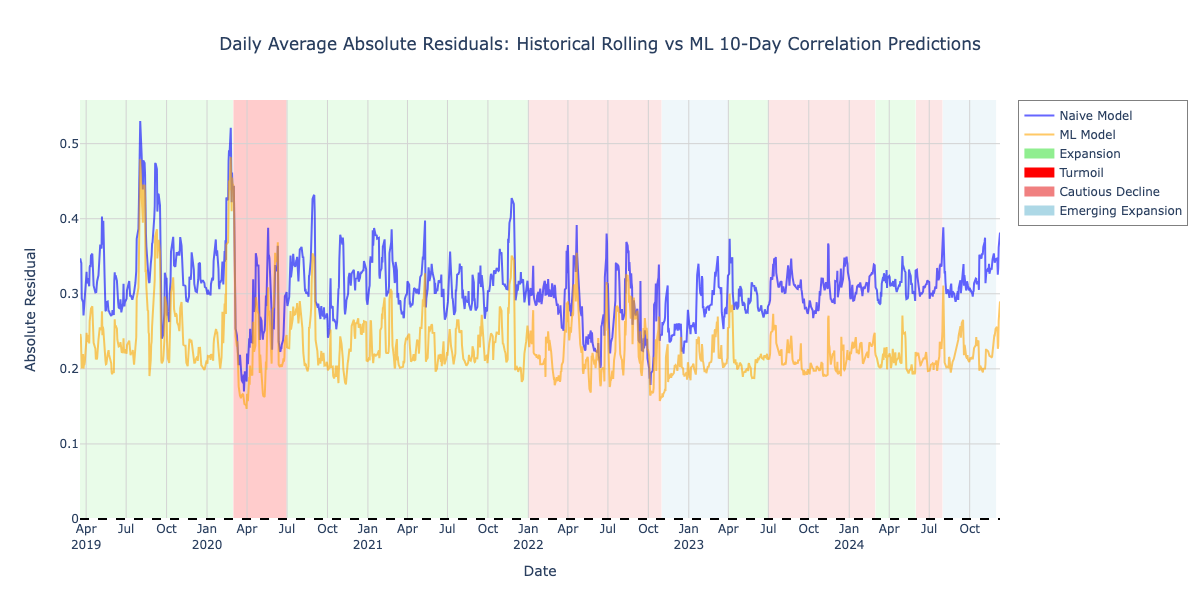} 
    \caption{Daily average absolute residuals between predicted and actual 10-day future stock-to-stock correlations  for a historical rolling window baseline and predicted THGNN correlations.}
    \label{fig:abs_ML} 
\end{figure}

Figures \ref{fig:abs_ML} depicts the absolute average residuals for both the models predicted correlation values and the naive historical correlation values respectively. The ML predicted correlations consistently has average absolute residuals of $0.20-0.30$ whereas, the naive predictions consistently predicts average absolute residuals of $0.30-0.40$. Importantly, the naive model residuals are higher throughout the whole period indicating that the ML model is able to increase predictive accuracy throughout varying market regimes. We also see that during periods of stress and uncertainty such as, COVID-19 (2020) and the Russia-Ukraine (Feb - July 2022) conflict, residuals spike for both the naive and predicted residuals. During these stress periods we see that the model has smaller absolute residuals then the naive baselines of approximately $0.05$ indicating that the model does adapt to crisis periods better than the historical baseline; although the model does still struggle to predict these inherently unpredictable periods with a higher absolute residual error than stability periods. 

We can further break down the residual analysis into the average absolute residuals per sector, Table \ref{tab:avg-abs-residual-by-sector-simple}. Average absolute residuals are tightly cluster, indicating the model's predictive accuracy generalizes broadly across sectors.

\begin{table}[H]
\caption{Average absolute residual between predicted and actual 10-day future stock-to-stock correlations for predicted THGNN correlations.}
\label{tab:avg-abs-residual-by-sector-simple}
\centering
\begin{tabular}{l r}
\toprule
\textbf{Sector} & \textbf{Average Residual} \\
\midrule
Real Estate                 & 0.298151 \\
Financials                  & 0.300287 \\
Materials                   & 0.302478 \\
Industrials                 & 0.303522 \\
Energy                      & 0.304135 \\
Consumer Discretionary      & 0.304656 \\
Information Technology      & 0.306587 \\
Utilities                   & 0.310401 \\
Communication Services      & 0.314644 \\
Health Care                 & 0.315660 \\
Consumer Staples            & 0.317321 \\
\bottomrule
\end{tabular}

\end{table}

\subsection{Baseline Replication and Benchmark Results}
\label{Replication}

To ensure comparability with existing statistical arbitrage frameworks, we implement the clustering an trading procedures describe in Korniejczuk and Ślepaczuk (2024). While we are able to closely match the reported non-ML baseline performance under comparable assumptions, we cannot reproduce the large performance gains reported for the ML-filtered strategy.

Classifier results, brier scores and precision, closely align with those reported in the original study, suggesting that differences arise at the portfolio construction and back testing stage rather than from implementation errors. Given this discrepancy, to avoid potentially unstable baselines trading performance is evaluated relative to the S\&P 500 buy-and-hold benchmark. See Appendix \ref{Replication1} detailed replication results.

\subsection{Trading Performance}

\begin{figure}[H] 
    \centering
    \includegraphics[width=1.0\textwidth]{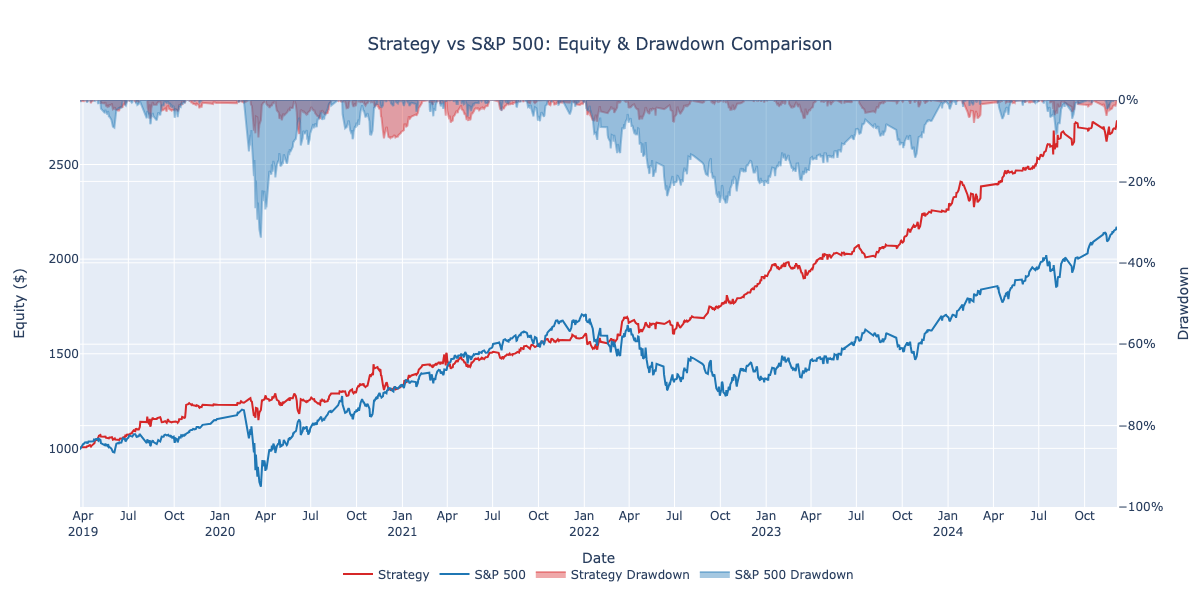} 
    \caption{Equity curve of forward-looking SPONGEsym basket trading strategy from 2019-2024 against the S\&P 500.}
    \label{fig:equity_curve} 
\end{figure}

The strategy generates strong out-of-sample performance when benchmarked against the S\&P 500 index between April $2019$ and October $2024$. Figure \ref{fig:equity_curve} plots the cumulative return of the strategy against the S\&P 500 with corresponding drawdowns.

The model exhibits a consistent upward trend with materially lower drawdowns than the benchmark, notably during the COVID-19 sell-off in early 2020. The strategy diverging from the benchmark during the Russia-Ukraine conflict in April $2022$. This is consistent with the model's ability to adapt correlation forecasts and basket construction under changing market regimes, while ML-based signal filtering mitigates losses during periods of elevated volatility.

Table \ref{tab:performance_comparison} summarizes key performance indicators for the strategy and the S\&P 500. The strategy achieves an annualized Sharpe ratio of $1.84$, substataially exceeding the S\&P 500's 0.75, while limiting maximum drawdown to $-9.4\%$ comparewd with $-33.9\%$ for the benchmark. These results indicate that the forward-looking correlation forecasts materially improve risk-adjusted performance relative to both historical clustering methods and passive buy-and-hold equity approaches. 

\begin{table}[H]
\centering
\caption{Performance metrics of the forward-looking SPONGEsym correlation basket trading strategy against the S\&P 500.}
\label{tab:performance_comparison}
\begin{tabular}{lcc}
\toprule
\textbf{Metric} & \textbf{Strategy} & \textbf{S\&P 500} \\
\midrule
Annualized Return (ARC) & 19.20\% & 14.43\% \\
Annualized Standard Deviation (ASD) & 0.121 & 0.204 \\
Sharpe Ratio & 1.837 & 0.647 \\
Sortino Ratio & 2.473 & 0.786 \\
Maximum Drawdown (MDD) & -9.43\% & -33.93\% \\
Maximum Loss Duration (MLD) & 82 days & 512 days \\
Calmar Ratio (CR) & 2.035 & 0.425 \\
Information Ratio (IR) & 0.133 & N/A \\
\bottomrule
\end{tabular}
\end{table}

To test if the Sharpe ratios of the strategy and the S\&P 500 differ significantly we employ the Ledoit-Wolf bootstrap test as the conditions for parametric tests are violated. A one-sided (greater than) Ledoit-Wolf bootstrap test is used as this preserves time-series dependence and provides robust inference for Sharpe ratios under non-normal distributions. Under the null-hypothesis that the Sharpe ratios are equal, a p-value of $0.0200$ allows rejection at conventional significance levels and indicates that the Sharpe ratio of the strategy is significantly greater than the S\&P's. Further, the bootstrap confidence intervals for the Sharpe difference suggests an economically meaningful, out performance. The $95\%$ percentile interval is $[0.054, 2.147]$ while the studentized $95\%$ interval (adjusts for heteroskedasticity) is $[-0.123, 2.074]$. Both intervals have all or predominant mass above zero respectively, indicating economically meaningful outperformance.

\subsection{Sector Attention}
A central question for the interpretability of our model is how attention weights, learned with the GAT, evolve across times and regimes. The daily-time series plots presented below capture the cumulative normalized average attention that the average stock in a sector $X$ gives to sector $Y$. In examining these plots we look for two key features. First, we expect to see persistent dependencies that align with known structural relationships, such as finance and real estate which tend to move together as the real estate market depends highly on bank lending and interest rates. Secondly, we anticipate regime-dependent shifts in attention, where the influence of certain sectors intensifies during market stress, crashes, or recovery. This allows us to confirm that the model is picking up on known economic trends and adjusting sector relationships under varying market conditions.

This analysis has both economic and trading applicability. From an economic perspective, shifts in sector-to-sector attention capture how the role of key sectors in transmitting shocks through the financial system evolves over time, particularly during crisis periods. From a trading perspective, these dynamics directly impact portfolio construction: through highlighting which sectors act as information hubs at different times, the attention weights can inform adaptive hedging ratios, basket design, and risk management system that respond to regime change rather than assuming static correlations or dependencies. 

Overall, the time-series plots of sector-to-sector importance serve not only as validation that the model is leveraging economically meaningful signals, but also as a tool for understanding how sector relationships change over time. In the main text, we focus on three sectors - Financials, Energy, and Real Estate - chosen for their economic significance and distinctive dynamics. Finance and Energy are among the largest sectors by number of stocks and market capitalization, while Real Estate is included given the notable patters observed in its dependencies. The remaining eight plots are provided in Appendix \ref{RemainingPlots}. The discussion first details the sector-level dynamics for each individual time series and then synthesizes the results into broader economic and trading implications.

\begin{figure}[H] 
    \centering
    \includegraphics[width=1.0\textwidth]{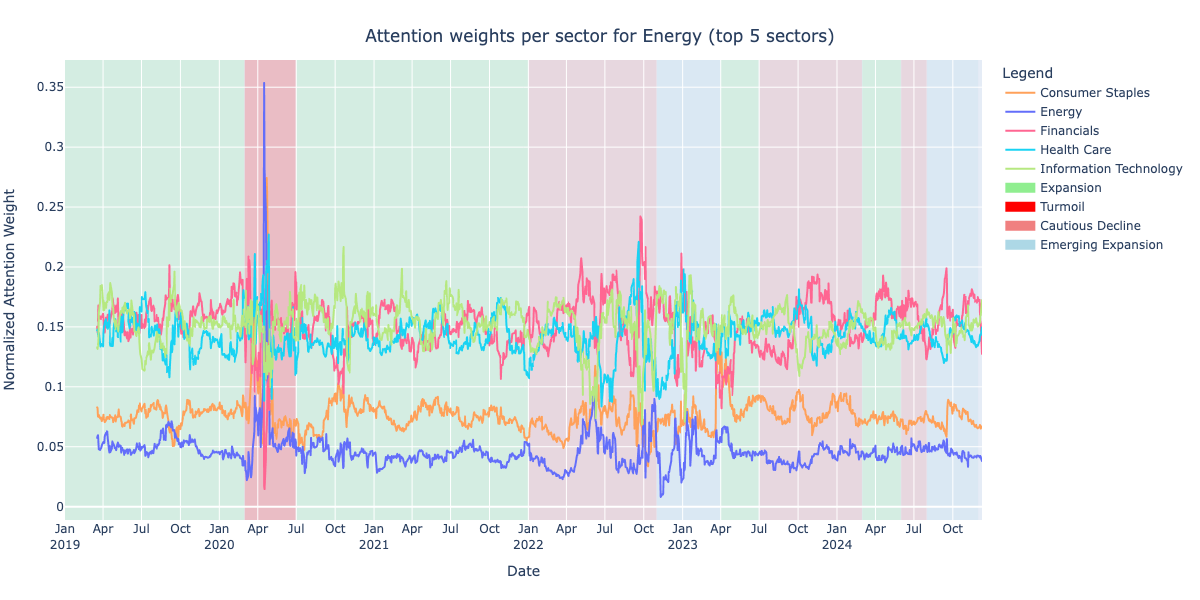} 
    \caption{Attention weights learned by the THGNN model for stocks in the Energy sector. Each data point shows the average attention an Energy stock gives to other sectors at a specific time, highlighting which sectors are the most important in forming predictions.}
    \label{fig:energy} 
\end{figure}

Figure \ref{fig:energy}, depicts the consistently high importance the Energy sector places on Financials, Health Care, and Information Technology, with Energy and Consumer Staples receiving consistently lower weightings. Financials, Health Care, and IT consistently cluster between $0.13-0.18$, with Energy and Consumer Staples consistently around $0.04$ - $0.09$. The high impact of Financials is intuitive given the highly competitive and capital-intensive nature of the energy industry. However, the reliance on IT and Health Care is less direct and likely represents some complex interdependence relationship.

A notable spike occurs during the COVID-19 turmoil period, where intra-sector Energy attention spikes from $0.05$ to $0.35$, suggesting heightened self-dependence when systemic volatility peaks. This aligns with the oil demand collapse induced by the conflict and the extreme price fall observed in April 2020, when the WTI futures contracts fell to $-37$ per barrel (CFTC, 2020). This shock is short-lived with attention weights that energy assigns to its top 5 sectors returning to normal levels quickly, illustrating the sector's sensitivity to commodity price shocks.

Overall, the model is largely economically intuitive and crash-type demand shocks ($2020$) push Energy to increase attention internally, while geopolitical supply shocks ($2022$) increase Energy's cross-sector reliance. 

\begin{figure}[H] 
    \centering
    \includegraphics[width=1.0\textwidth]{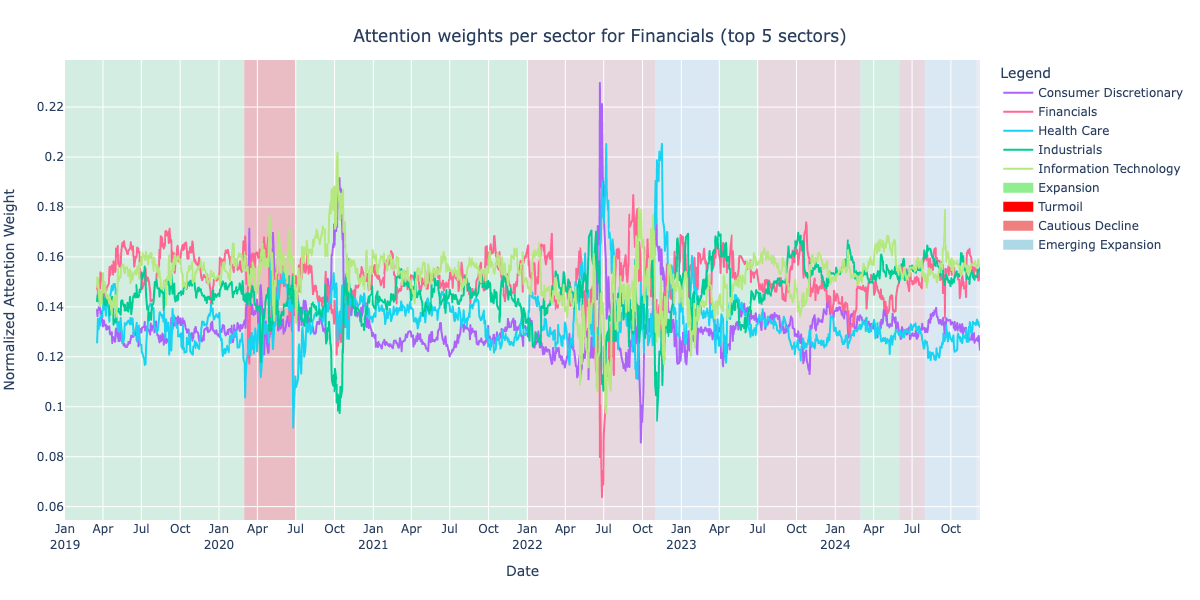} 
    \caption{Attention weights learned by the THGNN model for stocks in the Financials sector. Each data point shows the average attention a Financials stock gives to other sectors at a specific time, highlighting which sectors are the most important in forming predictions.}
    \label{fig:financials} 
\end{figure}

Figure \ref{fig:financials}, shows that Financials place consistent importance on Information Technology, Health Care, Consumer Discretionary, Industrials, and Financials. The attention in the industry is diversified with the top $5$ sectors consistently accounting for $75\%$ of total attention, with weights clustered between $0.12$ and $0.16$. This pattern is economically intuitive as the Financial sector underpins consumption, investment, and credit across the economy, and its linkages are broad rather than concentrated in a few sectors.

During the COVID-19 turmoil (January-July 2020), attention weights remain relatively stable indicating that Financials are influenced by system-wide shocks rather than any single sector. The stable allocation of the model here reflects the sector's exposure to system-wide effects rather than sector-specific dynamics. However, during October $2020$, a recovery phase marked by heightened volatility around U.S. elections, renewed COVID-19 outbreaks, and uncertainty over fiscal stimulus (Schroders, 2020), IT and consumer discretionary attention both spike to over 0.18. Unlike the first half of 2020, when central bank intervention dominated, this episode reflects idiosyncratic volatility within financial equities, with the model capturing a shift from systemic liquidity stress to sector-specific volatility. However, this effect was only for a short period of time indicating this idiosyncratic behavior was alleviated quickly by the well-diversified financial sector. 

During the period from February and October $2022$, corresponding to the Russia-Ukraine conflict and its spillovers through energy and commodity markets, attention weights again fluctuate. We observe increases in Consumer Discretionary and Health Care rather than in Energy or Financials, which seems counterintuitive. This likely reflects the model's recognition of indirect financial linkages. For example, higher energy prices transmit into inflation, which alters household consumption patterns in turn feeding into financial sector valuations and activities. The Financial sector is characterized by complex layered inter dependencies which do not have a simple mapping to economic intuition.

All together, the Financial plot illustrates the difficulty in attributing shocks to single drivers in sectors that have many complex inter dependencies. The sensitivity to macroeconomic policy, geopolitical tension, and consumption dynamics means that the model's attention distribution is broad and more volatile during episodes of sector-specific stress.

\begin{figure}[H] 
    \centering
    \includegraphics[width=1.0\textwidth]{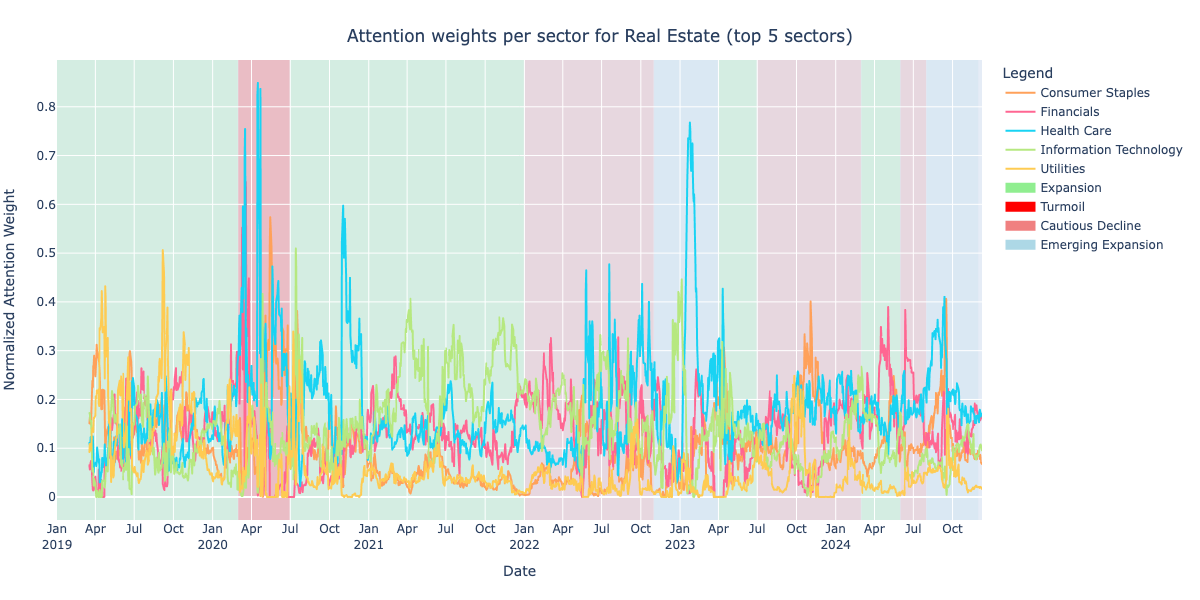} 
    \caption{Attention weights learned by the THGNN model for stocks in the Real Estate sector. Each data point shows the average attention a Real Estate stock gives to other sectors at a specific time, highlighting which sectors are the most important in forming predictions.}
    \label{fig:real_estate} 
\end{figure}

Figure \ref{fig:real_estate} stands out from the other sector plots as it is extreme volatile, the magnitude of its attention weights are large, and attention weights are not always directly intepretable or meaningful. Unlike other sectors, the weights do not cluster around a band and are have dramatic attention spikes exceeding $0.7$. These out sized fluctuations highlight the sector's complex and unstable interdependencies.

One of the most striking and persistent features is the consistently large weights assigned to Health Care. This relationship is very hard to quantify economically, with no natural huge sector linkages, other than perhaps health care construction projects or office spaces. Despite this, the broad volatility is economically meaningful. Real Estate sits at the intersection of many sectors: Financials (financing conditions), Industrials (construction activity) - surprisingly industrials is not in the top $5$ largest drivers however, as is revealed by Figure \ref{tab:avg-abs-residual-by-sector-simple} Industrials is the $5th$ largest average attention contributor -, Utilities (utilities and infrastructure) and Consumer Staples (consumer demand). The model's noisy nature reflects that the sector's performance is not tied to a single dependency rather, it depends on a wide range of fluctuating dependencies. In periods of stability we do see that Health Care does not have this huge weight as it does in unstable periods so these spikes are likely the model picking up on some complex interdependency.

Overall, the plot illustrates the promise and limitation of attention-based interpretability. The volatility confirms Real Estate's systemic sensitivity however, the dominance of Health Care attention suggests that not all outputs can be mapped to intuitive sectoral linkages. From a trading perspective, Real Estate's economically uninterpretable attention spikes provide limited signal generation but could be used as indicators of regime shifts where volatility propagates across markets.

\begin{figure}[H] 
    \centering
    \includegraphics[width=1.0\textwidth]{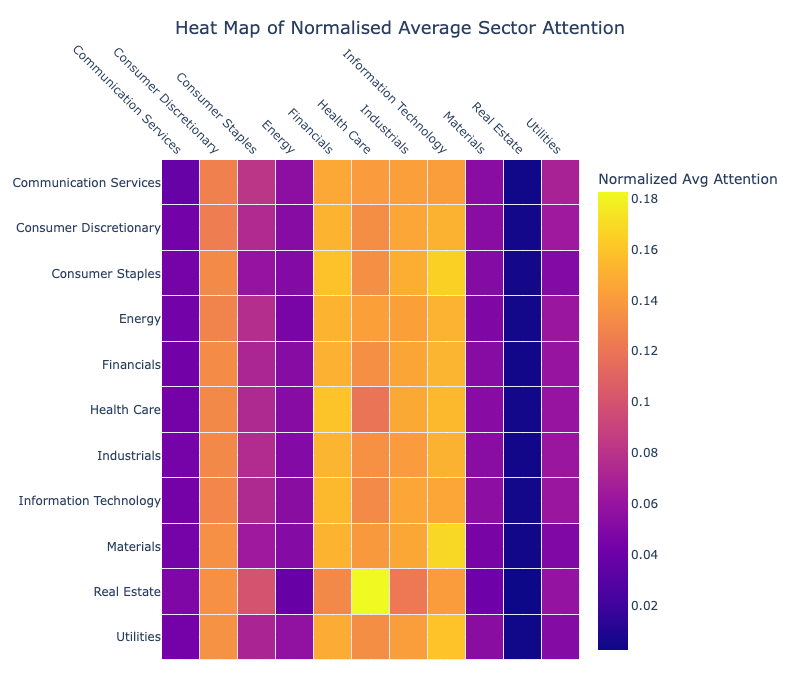} 
    \caption{Heatmap of average attention weights learned by the THGNN model for all 11 sectors. Each cell represents the average attention weight the model assigns to a sector when producing correlation predictions for stocks in that sector.}
    \label{fig:Avg_sector_attn} 
\end{figure}

Figure \ref{fig:Avg_sector_attn} presents the heat map of normalized average sector attention weights across the whole out-of-sample period. This visualization provides a summary of the long-run structural dependencies learned by the model. 

The most striking pattern is the vertical influence relationship: sectors that are important for one sector tend to be important for others too, while sectors with low influence contribute little across the board. Part of this is mechanical -  sectors with stocks naturally contribute a larger cumulative share - however, the model naturally weights individual stocks by importance therefore this is a real relationship. The pattern suggests that the dynamics of the market and stock relationships can be explained through a handful of sectors that capture both complex inter-dependencies and intuitive economic relationship. Contrastingly, sectors such as Real Estate are not influential as measured by its attention weights, however this is not to say they lack importance because their relationships with other sectors may be captured indirectly through other dominant sectors.

Secondly, despite Energy consistently appearing on plots due to its large spikes during uncertain periods, energy receives low average weight. This indicates that Energy's influence is highly regime-dependent; it is important during crisis, when oil prices or commodity supply shocks dominate, but muted in normal conditions when other drivers better explain market correlations. 

\subsection{Feature Importance}

\begin{figure}[H] 
    \centering
    \includegraphics[width=1.0\textwidth]{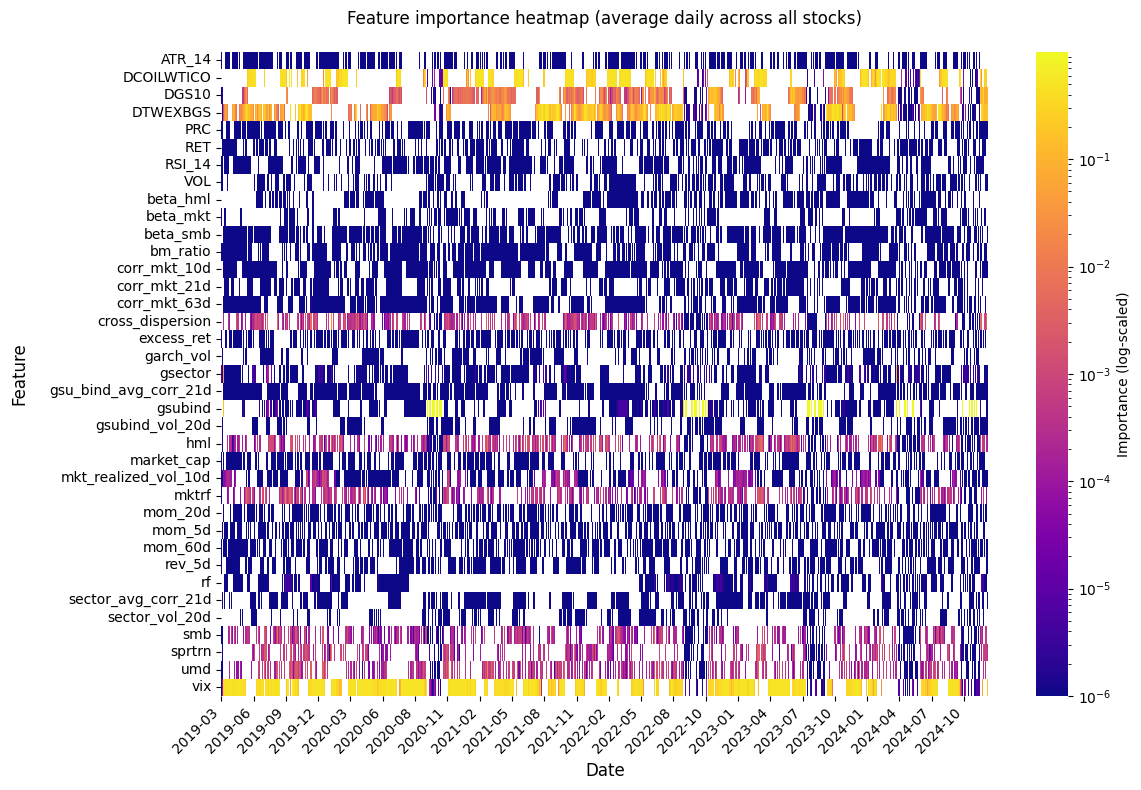} 
    \caption{Heatmap of feature importance weightings given by the transformers to all 37 input features in building the latent stock embeddings. Each cell shows the average normalized importance of a feature.}
    \label{fig:Heat_all_features} 
\end{figure}

Figure \ref{fig:Heat_all_features} depicts a heatmap of the importance weights, Gradient $\times$ Input, by the transformer to features when building stock embeddings.

The striking result in Figure \ref{fig:Heat_all_features} is that, despite the large number of candidate features, only a handful dominate the learned embeddings. This suggests that the model's correlation predictions are highly sensitive to movements in a small subset of variables, while many other contribute very little. This indicates that the transformer focuses on the key drivers of correlation dynamics rather than uniform weighting.

\begin{figure}[H] 
    \centering
    \includegraphics[width=1.0\textwidth]{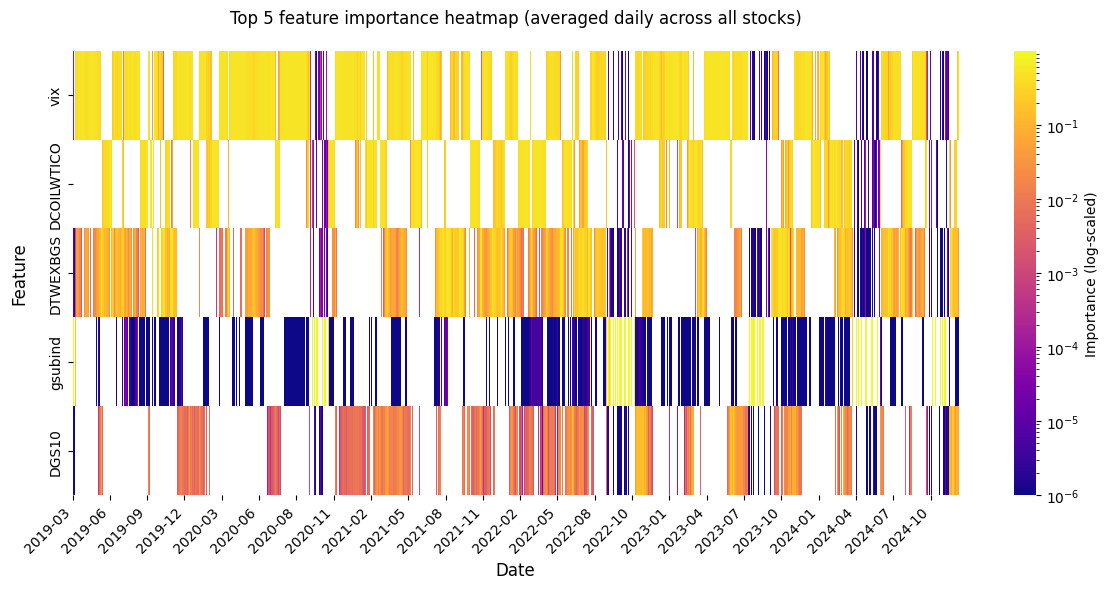} 
    \caption{Heatmap of the top 5 highest feature importance weightings given by the transformers to features in building the latent node embeddings. Each cells shows the average normalized importance of a feature.}
    \label{fig:Heat_top5_features} 
\end{figure}

Figure \ref{fig:Heat_top5_features} depicts a heat map of the five most important features: the VIX, crude oil prices (DCOILWTICO), U.S. 10-year Treasury yields (DGS10), trade-weighted broad dollar index (DTWEXBGS), and sector binding (gsubind).

The prominence of the VIX is intuitive, reflecting volatility as a key driver of correlation spikes during crisis and stable periods. Crude oil, treasury yields, and the dollar index reflect the macroeconomic state of the economy as well as commodity price movements in shaping correlation predictions. The role of gsubind is particularly notable. During periods of stability gsubind plays a limited role however, in periods of crisis such as, COVID recovery in November $2020$, gsubind becomes particularly important. In these times, reassigning a stock to a different sector would substantially alter its correlation predictions. Contrastingly, during stable periods macro factors dominate, indicating that stock embeddings are not purely sector-dependent but also capture global market conditions.
 
These results indicate that correlation dynamics are concentrated around a handful of macro drivers and structural features rather than being evenly spread across inputs.

\section{Conclusion and Discussion}

This study replaces backward-looking rolling correlation estimates with forward-looking, regime-aware predictions and tests whether they improve basket construction, trading performance, and our understanding of what drives equity clustering during periods of crisis. Rolling-window correlations are slow to adapt to market regime shifts therefore, when regimes shift they generate large predictive residuals and losses in trading strategies. Current literature generates correlation predictions on short horizons which generates limited use outside of high-frequency trading. Therefore, we implement a novel Temporal-Heterogeneous GNN (THGNN) that combines a Transformer temporal encoder with a graph-attention module over the equity network allowing for both temporal and cross-asset structure to be considered. The prediction task was framed as a 10-day stock-stock correlation forecasting as residual learning in Fisher-$z$ space with a baseline of the past 20-day rolling window correlations. Predictions from the model are passed into the SPONGEsym clustering algorithm and filtering through an ML ensemble to construct market-neutral basket trades.

The out-of-sample period (2019-2024) showed the model is able to reduce MAE from 0.3071 (20-day historical rolling window) to 0.2302 and RMSE from 0.3852 to 0.2940. The model also reduced bias and improves both Pearson and Spearman correlation more than double to 0.778 and 0.795 respectively. The improvement in predictive accuracy over the historical baseline persists across varying market regimes such as, COVID-19 and Russia-Ukraine conflict, however significant increases in residuals do occur during periods of crisis and market turmoil. Residuals for all sectors remain tightly clustered suggesting the model generalizes to all equity types in the S\&P 500. Therefore, we can conclude that a transformer and GNN ML architecture can meaningfully increase predictive accuracy over historical rolling window baselines but still generates large residuals during crisis periods.

Attention diagnostics show that the model consistently emphasizes a small set of macro-financial drivers - most notably market volatility (VIX), crude oil prices, the 10-year U.S. Treasury yield, the trade-weighted dollar, and sector identifiers - with these weights rebalancing across market regimes. This aligns with economic intuition, as volatility, interest-rate, and commodity channels are central to shifts in equity correlation structure. At the network level, graph attention reveals clear regime dependence in sectoral relationships, with Energy exhibiting heightened intra-sector dependence during COVID-19 and broader cross-sector linkages emerging during the Russia–Ukraine conflict. While some attention reallocations lack direct economic interpretation, this reflects the nonlinear and high-dimensional nature of the learned dependency structure rather than model instability. Taken together, the joint evolution of temporal and relational attention indicates that the drivers of equity correlations are regime-dependent and unstable over time, supporting the use of forward-looking, adaptive models in place of static rolling-window estimates.

When integrated into SPONGEsym clustering and filtered by the ML ensemble, the strategy generates 19.20\% annualized return with a Sharpe of 1.837, Maximum Drawdown (MDD) -9.43\%, and Maximum Loss Duration (MLD) 82 days from 2019-2024 versus the S\&P 500's 14.43\%, 0.647, -33.93\%, and 512 days respectively. Therefore, the results indicate that the model is able to improve both returns and risk-adjusted returns compared to the S\&P 500. 

For quant firms and traders, the results represent a wider shift. Variables like correlation should be treated as predictable variables not fixed summaries of the past. Correlations are not long-run stable economic measures rather, their drivers adjust with regime shifts and a deep learning architecture can capture the complex noisy inter-dependencies of short-medium horizon correlations. Strategies should be based off predicted state variables not fixed summaries of the past. Specifically, this study shows that in a basket trading setting rolling windows should be swapped for forward predictions, models should be fed macro variables to ensure regime adaptability, and hedging should be informed by future predictions. 

It is important to acknowledge the limitations of this study. Firstly, no slippage costs are considered in the backtest so realized performance may be overstated. Secondly, the THGNN has not been compared to alternative ML models (LSTM, CNN, alternative GNN-based models), limiting comparative claims. Thirdly, the model has been trained only under one set of hyperparameters (such as Transformer sequence length, learning rate, attention heads), and therefore it is almost certain predictive accuracy is sub-optimal. Lastly, interpretability relies on one family of attribution tools, limiting the robustness of the interpretability findings and potentially their depth. Moreover, the substantial gap between the replication results and those reported by Korniejczuk and Slepaczuk (2024) highlights the fragility of backtests, the need to critically evaluate reported results, and underscores the need for cautious interpretation of historical performance. Despite these limitations, the finding that forward-looking correlation estimates can improve basket construction in this setting is still valid, but they do constrain the strength and generality of our claims.

Future work should explore whether alternative architectures, such as LSTMs, CNNs, or alternative GNN variants, can further increase predictive accuracy, thereby clarifying whether the THGNN is uniquely well-suited to this task as theory suggests. Extending the analysis to other equity universes (global equities) or asset classes (cryptocurrencies) would test the external validity of the approach. On the trading side, deep models can be used as classifiers for signal filtering which may improve their predictive accuracy thereby boosting profits. Finally, combining attention with attribution tools such as Integrated Gradients, DeepLIFT, SHAP, or Gradient $\times$ Input, could provide more robust insight into the economic drivers of predicted correlations.

\newpage
\section{Reference List}
\nocite{*}
\printbibliography[heading=none]

\appendix
\section{Appendix}
\subsection{Optimization and training}
\label{Training}
We use AdamW with learning rate $3\times 10^{-4}$, betas=$0.9/0.999$, weight decay=$2\times 10^{-4}$ with bias/LayerNorm excluded from decay. AdawW is well suited to this setting as it combines adaptive step sizes - important given gradients vary across features and regimes - with decoupled weight decay, which regularizes the model without adjusting the effective learning rate. The learning rate falls within the typical range for Transformer's and allows the model to learn quickly without overshooting in noisy, high-variance regions of the loss surface. The small weight ensures that informative information from the noisy correlation structure is not over regulated, whilst curbing overfitting. Bias and LayerNorm are excluded from decay as they set basic scales and shrinking them led to poor calibration. 

The learning rate follows a cosine decay schedule updated each optimizer step with $T_{\max} =$ total optimizer steps and a minimum learning rate of $1\times 10^{-6}$. Cosine decay provides fast initial exploration followed by a smooth annealing that switches the model focus to fine-tuning after it has discovered the coarse structure of the correlation surface. This avoids oscillations that arise from poorly tuned decay, which is very important for non-stationary financial data where the loss landscape evolves over time. 

Given that graphs are large, a batch size of 3 with gradient accumulation of size 6 is used giving an effective batch size of 18. This allows the model to fit on GPU memory and ensures the batch is large enough to average idiosyncratic noise in individual days, whilst keeping training responsive to new information. Gradients are clipped at 1.0 to stop shocks from blowing up an update and affecting convergence. Standard Pre-norm with LayerNorm is applied before attention and feed-forward blocks. This ensures that activation is in a stable range and improves gradient flow through multiple layers, critical for learning long-horizon dependencies. All weights are initialized with Xavier-uniform and zero biases giving the model a neutral, balanced starting point. The model is trained over 75 epochs to ensure enough time for the model to reach convergence. Beyond 75 epochs, the model will be overfitting, such that additional training yields only marginal increases, if any. Overall, these training parameters ensure the model adapts quickly to regime changes without overreacting to noise.

\subsection{ML Filtering Training}
\label{MLFiltering}
The models are trained under the feature set proposed by Korniejczuk and Slepaczuk (2024) with features extracted for every trading day - such that, after a 10-day hold period we move forward one day not 10 to allow more data points to be collected. Specifically, the ML classifiers are trained on the following features:

\textbf{Local vertex degree:}  

Measures the connectivity of a node within its own sub-graph.  
\[
d^{\text{local}}_i = \frac{1}{S-1} \sum_{n=1}^{S} e_{i,n}, 
\]
where $e_{i,n}$ is the weight of the $n$-th edge attached to vertex $i$, and $S$ is the size of the sub-graph.  

\textbf{Global vertex degree:}  

Normalizes the degree of a node with respect to the entire graph:  
\[
d^{\text{global}}_i = \frac{1}{G-1} \sum_{n=1}^{G} e_{i,n}, 
\]
where $G$ denotes the total number of vertices in the graph.  

\textbf{Graph density:}  

Captures how densely connected a sub-graph is:  
\[
\delta = \frac{1}{S(S-1)} \sum_{i=1}^{S} \sum_{n=1}^{S} e_{i,n}, 
\]
with $S$ again denoting the sub-graph size.  

\textbf{Cluster size:}  

Represents the proportion of nodes in cluster $j$:  
\[
c_j = \frac{|v_j|}{G},
\]
where $|v_j|$ is the number of vertices in cluster $j$, and $G$ is the total graph size.  

All graph-based measures are normalized either by cluster size (for within-cluster comparability) or by global graph size (to account for variations in the number of available stocks, which may not always equal 500).  

We also extract the following return features:
\begin{itemize}
    \item Cumulative returns deviation from the cluster average over the last five days
    \item Direction of deviation relative to the cluster mean (long/short indicator)
    \item Mean cluster return for the last 10 days
    \item Mean stock return for the last 10 days
\end{itemize}

We choose not to add more features to this dataset, such as the 37 features that the THGNN is trained on. This is because models classifiers performance may degrade when confronted with high-dimensional or weakly informative inputs (Guyon \& Elisseeff, 2003). Hyperparameters for each classifier and each feature set are selected through a grid search, the tables below summarizing the grid searches, with bold values representing the best performing mix of parameters. 

\begin{table}[H]
\centering
\caption{Hyperparameter grid search result for MLP Classifier trained to predict the profitability of a trading signal given return and graph based features. }
\label{tab:mlp}
\small
\begin{tabular}{llllll}
\toprule
\textbf{Hidden Layer} & \textbf{Activation} & \textbf{Alpha} & \textbf{Learning} & \textbf{Batch} & \textbf{Solver} \\
\textbf{Sizes} & \textbf{Function} & & \textbf{Rate} & \textbf{Size} & \\
\midrule
64,64 & \textbf{tanh} & 0.1 & constant & \textbf{50} & \textbf{sgd} \\
64,32 & relu & 0.01 & \textbf{adaptive} & 200 & adam \\
\textbf{64} & sigmoid & 0.0001 & & & \\
32,16 & & 0.000001 & & & \\
 & & \textbf{0.0000001} & & & \\
 & & 0.00000001 & & & \\
\bottomrule
\end{tabular}
\begin{flushleft}
\footnotesize Note: Bold values indicate the optimal hyper-parameters selected through grid search.
\end{flushleft}
\end{table}

\begin{table}[h]
\centering
\caption{Hyperparameter grid search result for AdaBoost Classifier trained to predict the profitability of a trading signal given return and graph based features.}
\label{tab:adaboost}
\begin{tabular}{ll}
\toprule
Number of estimators & Learning rate \\
\midrule
3   & 0.001 \\
5   & 0.01 \\
10  & 0.1 \\
20  & \textbf{1} \\
50  & 10 \\
\textbf{100} & \\
\bottomrule
\end{tabular}
\begin{flushleft}
\footnotesize Note: Bold values indicate the optimal hyper-parameters selected through grid search.
\end{flushleft}
\end{table}

\begin{table}[H]
\centering
\caption{Hyperparameter grid search result for HGB Classifier trained to predict the profitability of a trading signal given return and graph based features.}
\label{tab:hgb}
\begin{tabular}{llll}
\toprule
Learning rate & Early stopping & Max iter & Warm start \\
\midrule
0.01 & \textbf{auto}   & 50  & False \\
\textbf{0.1} & False  & \textbf{100} & \textbf{True} \\
0.5  &       & 150 & \\
1    &       & 250 & \\
5    &       & 500 & \\
\bottomrule
\end{tabular}
\begin{flushleft}
\footnotesize Note: Bold values indicate the optimal hyper-parameters selected through grid search.
\end{flushleft}
\end{table}

\begin{table}[H]
\centering
\caption{Hyperparameter grid search result for SGD Classifier trained to predict the profitability of a trading signal given return and graph based features.}
\label{tab:sgd}
\begin{tabular}{llllll}
\toprule
Loss function  & Penalty & Alpha   & Max iter & Early stopping & Learning rate \quad Warm start \\
\midrule
modified Huber & \textbf{l2}      & 0.5     & 200      & True              & constant \quad 1 \\
\textbf{log loss}       & l1      & 0.1     & \textbf{1000}     & \textbf{False}              & \textbf{optimal} \\
hinge          & elastic net & 0.01   & 10000    &                & adaptive \\
               &         & 0.001 &          &                & invscaling \\
               &         & 0.0001   &          &                & \\
\bottomrule
\end{tabular}
\begin{flushleft}
\footnotesize Note: Bold values indicate the optimal hyper-parameters selected through grid search.
\end{flushleft}
\end{table}

\begin{table}[H]
\centering
\caption{Hyperparameter grid search result for Logistic Regression trained to predict the profitability of a trading signal given return and graph based features.}
\label{tab:logreg}
\begin{tabular}{llllll}
\toprule
C   & Penalty & Solver  & Max iter & Class weight & Warm start \\
\midrule
0.1 & \textbf{l2}      & lbfgs   & 50       & Balanced & \textbf{True} \\
0.5 & l1      & liblinear & 75     & \textbf{None}        & False \\
\textbf{1}   & elastic net & \textbf{newton-cholesky} & 100 & & \\
1.5 &         &         & 120      &             & \\
2   &         &         & \textbf{200}      &             & \\
5   &         &         & 300      &             & \\
8   &         &         & 400      &             & \\
10  &         &         & 500      &             & \\
12  &         &         &          &             & \\
15  &         &         &          &             & \\
20  &         &         &          &             & \\
\bottomrule
\end{tabular}
\begin{flushleft}
\footnotesize Note: Bold values indicate the optimal hyper-parameters selected through grid search.
\end{flushleft}
\end{table}

\subsection{Sharpe Ratio Significance Tests}
\label{Significance}
To determine if the returns have autocorrelation a Ljung-Box test is conducted (Ljung \& Box, 1978). Specifically, the null hypothesis of residuals having no autocorrelation is tested through calculating the Q-statistic,
\[Q = n(n+2)\sum_{k=1}^h \frac{p_k^2}{n-k},\]
where n is the sample size, h is the number of lags being tested, and $p_k$ is the sample  autocorrelation at lag k.

To test for conditional heteroskedasticity, am ARCH LM test is used by regressing the squared residual on their own lagged square residuals to determine if today's squared return is predictable from past squared returns, which indicates time-varying variance (Engle, 1982). Rejection of the null hypothesis means there is evidence of heteroskedasticity. The LM statistic is calculated as,
\[LM = TR^2,\]
where $R^2$ is the regression $R^2$ and T is the sample size.

Non-normality (heavy tails) is tested through the Shapiro-Wilk test which compares the empirical order statistics to what you would expect under a normal distribution (Shaprio \& Wilk, 1965). Let $x_1\leq\dots\leq x_n$ be the ordered sample, with sample mean $\bar{x}$. The Shapiro-Wilk statistic is,
\[W = \frac{(\sum_{i=1}^{n} a_i x_i)^2}{\sum_{i=1}^n (x_i - \bar{x})^2}\]
where the weights $a_i$ are fixed constants computed from the expected order statistic of a standard normal. Under the null-hypothesis of normality, W is close to 1. A small W with small p-value indicates departures from normality.

To test if the Sharpe ratios are significantly different bootstrapping can be used as this requires no assumptions about parametric returns. We employ the Stationary Bootstrap (SB) to preserve serial dependence while avoiding sensitivity to a fixed block length (Politis \& Romano, 1994). SB draws random-length block with geometric distribution $L \sim \text{Geom}(p)$, so the mean block length is therefore $l= 1/p$ (Politis \& White, 2004). Blocks are concatenated until a bootstrap of length $T$ is formed and $B=5000$ separate bootstraps are formed. For each replication, $b=1, \dots, B$, $\Delta S^b$ is calculated and we can therefore form a confidence interval (Politis 
\& White, 2004). Let $\{\Delta S ^b\}^B_{b=1}$ be the bootstrap sample, the two sided ($1-\alpha)$ percentile CI is
\[q_{\alpha/2}(\Delta S), q_{1-\alpha /2}(\Delta S)\],
where $q_p(\dot)$ denotes the p-quantile of the bootstrap.

To improve coverage under heteroskedasticity and heavy tails, a studentized CI is also computed (Politis \& White, 2004). For each bootstrap, b, a T statistic is calculated as,
\[T^b = \frac{\Delta S^b - \Delta S}{se^b} = \frac{\Delta S^b - 0}{se_b},\]
where se is the standard-error estimate for $\Delta S$ from the original sample, and $\Delta S=0$ as this is the null hypothesis.
The CI can then be computed as,
\[\Delta S - T_{1-\alpha /2} se, \Delta S - T_{\alpha /2} se,\]
where $T_z$ is the quantile of $\{T^b\}_{b=1}^{B}$.
Using the studentized statistic, the observed value is $T_obs = \Delta S /se$ and the p-value is therefore computed from $\{T^b\}$,
\[p_{greater} = \frac{1}{B}\sum_{b=1}^B 1\{T^b \geq T_{obs}\}\]
\[p_{two-sided} = \frac{1}{B}\sum_{b=1}^B 1\{|T^b| \geq |T_{obs}|\}\]
Any $T_{obs}$ that meets our condition is set to 1 and any that doesn't is set to 0, the p-value is then the total percentage of $b=1,\dots,B$ that become 1.

\subsection{Replication Results}
\label{Replication1}
We first replicate Korniejczuk and Ślepaczuk's (2024) non-ML variant of the trading strategy. Following their methodology, we construct a 60-day rolling correlation matrix of residual returns and treat this as the signed, weighted adjacency matrix on which the SPONGEsym algorithm is applied. Clusters are formed form these historical residual-return correlations, contrarian weights are generated within each cluster, and the portfolio is rebalance every three days. Table \ref{tab:replication-trading} reports performance both without transaction costs and with a cost of $0.05\%$ per trade from 2000-2024. 

\begin{table}[H]
\caption{Trading performance for the author’s implementation (OWN) and the original strategy with and without transaction costs (TC), and SPY benchmark.}
\label{tab:replication-trading}
\centering
\begin{tabular}{lrrr}
\toprule
\textbf{Strategy} & \textbf{AR (\%)} & \textbf{Sharpe} \\
\midrule
Own (No TC)          & 8.27  & 0.86 \\
Own (0.05\% TC)      & -0.46 & 0.00 \\
Original (No TC)     & 10.24 & 1.17 \\
Original (0.05\% TC) & 2.44 & 0.28 \\
SPY (Benchmark)      & 6.48  & 0.33 \\
\bottomrule
\end{tabular}
\end{table}

The generated results are similar with our implementation generating slightly lower returns. This is likely due to the fact that we use data from WRDS (CRSP) whereas, Korniejczuk and Ślepaczuk (2024) use data from Yahoo Finance. Yahoo's historical files exclude delisted securities, therefore introducing survivorship bias by omitting names that experience rapid declines prior to delisting. Contrastingly, WRDS retains return histories for delisted firms therefore delisted stocks that would generate fast negative returns are considered in our implementation but not from Yahoo Finance data (as Yahoo Finance no longer stores information for desisted stocks).

Now, we implement the ML method described by Korniejczuk and Ślepaczuk (2024). Here we set the rebalance period to 10 days, generate the feature set described in the methodology, and train the ML ensemble over the same grid-search described in the methodology. The trained ML models are then used in a live backtest to filter positions with trades only undertaken if the probability of profitability is greater than the $90\%$ threshold.

Table \ref{tab:replication_classifiers} shows the Brier scores and precision of the trained ML classifiers for both our implementation and Korniejczuk and Ślepaczuk (2024). 

\begin{table}[H]
\centering
\caption{Comparison of Original and Replication Classifier Performance for ensemble of ML models trained to predict the profitability of a trading signal.}
\label{tab:replication_classifiers}
\small
\begin{tabular}{lcccc}
\hline
\textbf{Classifier} & \textbf{Original} & \textbf{Original} & \textbf{Replication} & \textbf{Replication} \\
 & \textbf{Brier Score} & \textbf{Precision} & \textbf{Brier Score} & \textbf{Precision} \\
\hline
MLP & 0.243 & 0.568 & 0.2423 & 0.5840 \\
ADA Boost & 0.247 & 0.544 & 0.2450 & 0.5454 \\
HistGradientBoosting & 0.218 & 0.653 & 0.2232 & 0.6452 \\
SGD & 0.247 & 0.547 & 0.2427 & 0.5481 \\
Logistic Regression & 0.249 & 0.580 & 0.2424 & 0.5840 \\
\hline
\end{tabular}
\end{table}

Table \ref{tab:replication_classifiers} shows that we are able to generate similar Brier scores and precision as Korniejczuk and Ślepaczuk (2024). Our ensemble - with the HGB model receiving double the weight of other models in the weighted soft-voting ensemble due to its performance - generates a $90\%$ profitability probability threshold of $0.6072$ compared to the papers threshold of $0.602$. 

\subsection{Remaining Sector Attention Plots}
\label{RemainingPlots}

\begin{figure}[H] 
    \centering
    \includegraphics[width=1.0\textwidth]{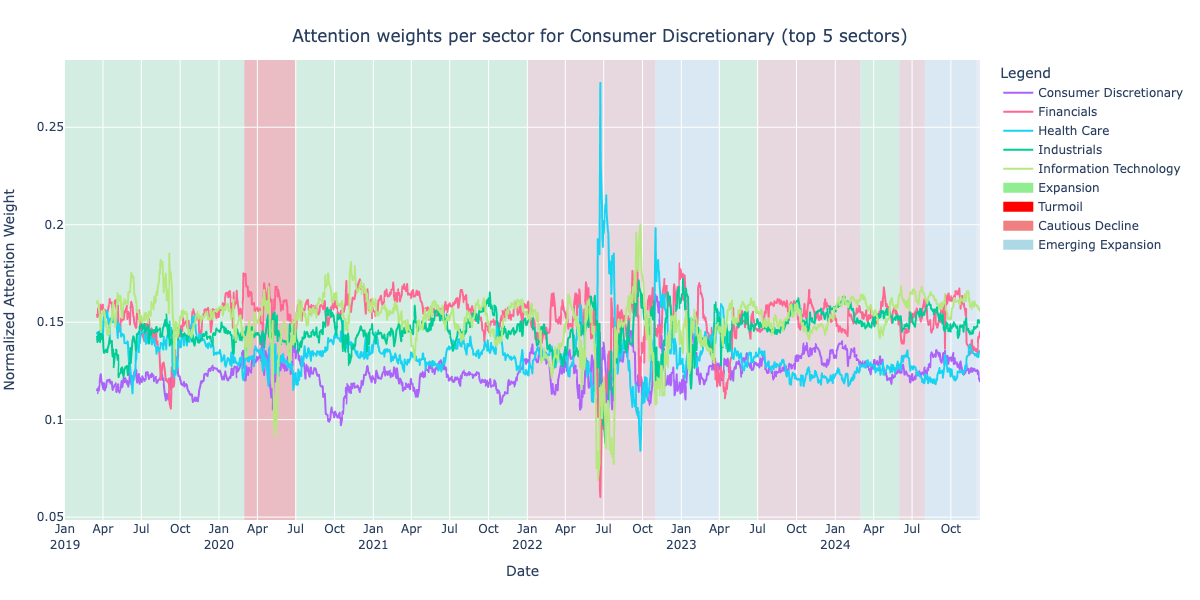} 
    \caption{Attention weights learned by the THGNN model for stocks in the Consumer Discretionary sector. Each data point shows the average attention a Consumer Discretionary stock gives to other sectors at a specific time, highlighting which sectors are the most important in forming predictions.}
    \label{fig:consumer_discretionary} 
\end{figure}

\begin{figure}[H] 
    \centering
    \includegraphics[width=1.0\textwidth]{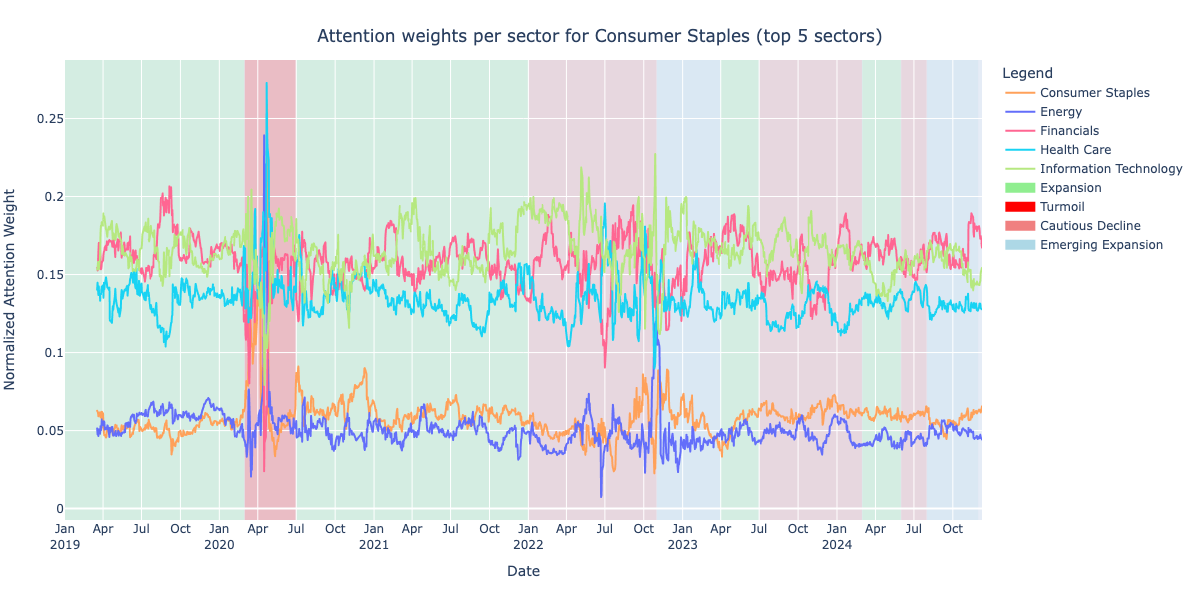} 
    \caption{Attention weights learned by the THGNN model for stocks in the Consumer Staples sector. Each data point shows the average attention a Consumer Staples stock gives to other sectors at a specific time, highlighting which sectors are the most important in forming predictions.}
    \label{fig:consumer_staples} 
\end{figure}

\begin{figure}[H] 
    \centering
    \includegraphics[width=1.0\textwidth]{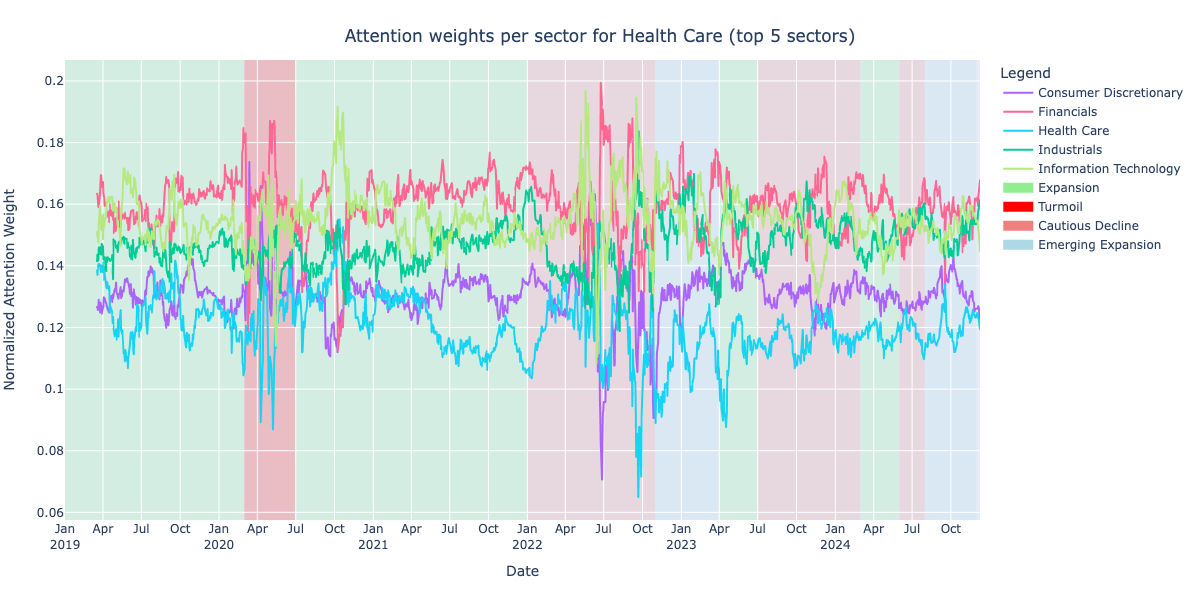} 
    \caption{Attention weights learned by the THGNN model for stocks in the Health Care sector. Each data point shows the average attention a Health Care stock gives to other sectors at a specific time, highlighting which sectors are the most important in forming predictions.}
    \label{fig:health} 
\end{figure}

\begin{figure}[H] 
    \centering
    \includegraphics[width=1.0\textwidth]{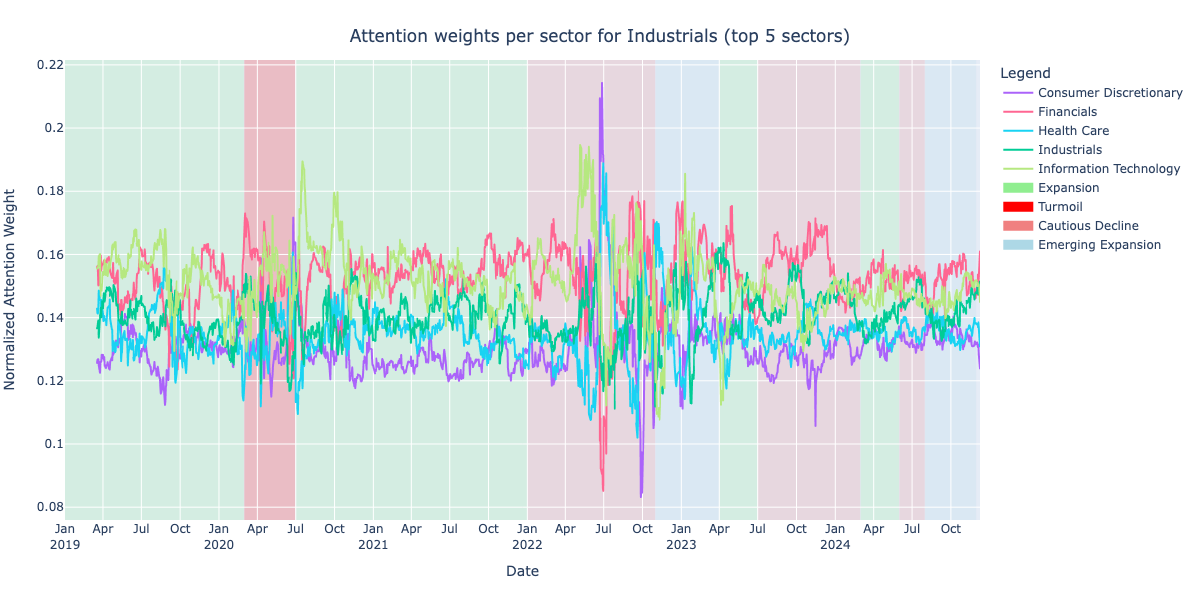} 
    \caption{Attention weights learned by the THGNN model for stocks in the Industrials sector. Each data point shows the average attention an Industrials stock gives to other sectors at a specific time, highlighting which sectors are the most important in forming predictions.}
    \label{fig:Industrials} 
\end{figure}

\begin{figure}[H] 
    \centering
    \includegraphics[width=1.0\textwidth]{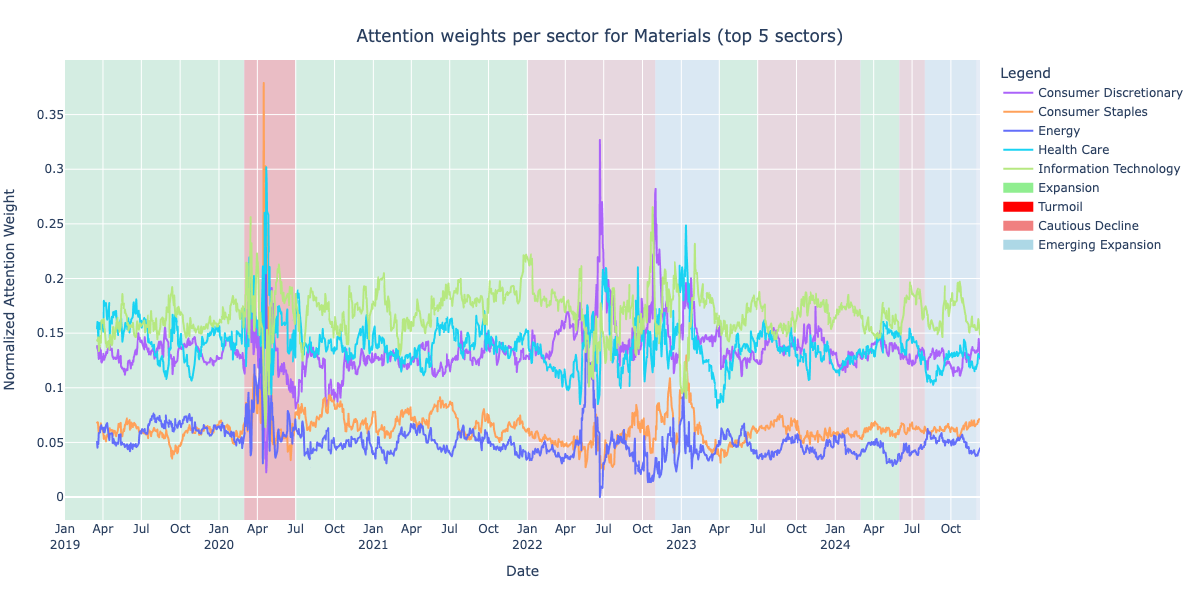} 
    \caption{Attention weights learned by the THGNN model for stocks in the Materials sector. Each data point shows the average attention a Materials stock gives to other sectors at a specific time, highlighting which sectors are the most important in forming predictions.}
    \label{fig:Materials} 
\end{figure}

\begin{figure}[H] 
    \centering
    \includegraphics[width=1.0\textwidth]{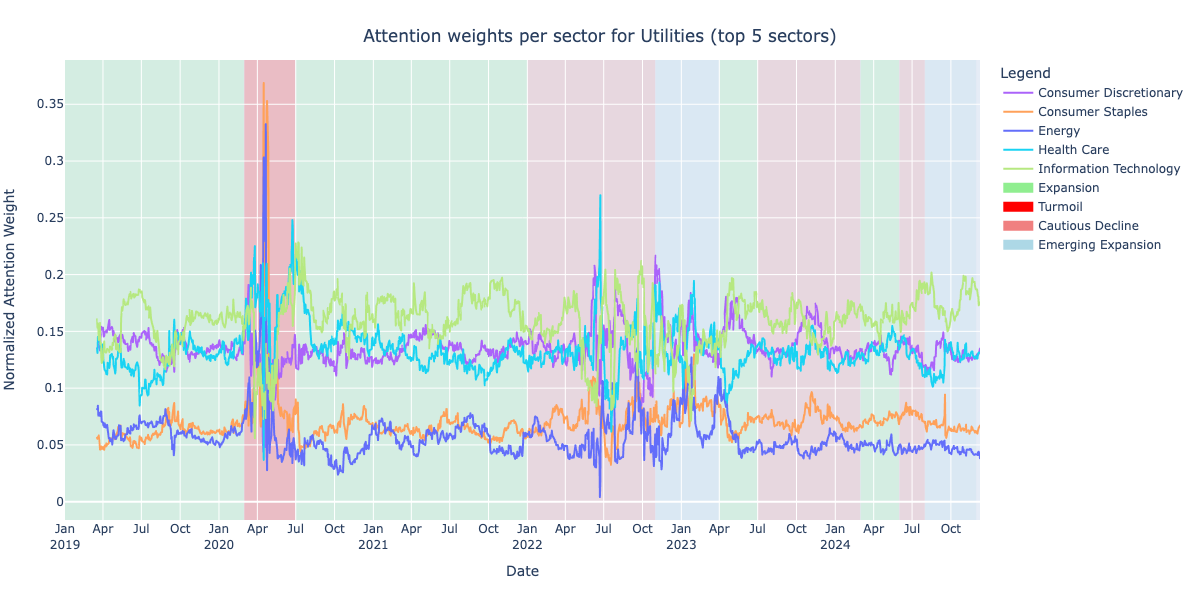} 
    \caption{Attention weights learned by the THGNN model for stocks in the Utilities sector. Each data point shows the average attention a Utilities stock gives to other sectors at a specific time, highlighting which sectors are the most important in forming predictions.}
    \label{fig:Utilities} 
\end{figure}

\begin{figure}[H] 
    \centering
    \includegraphics[width=1.0\textwidth]{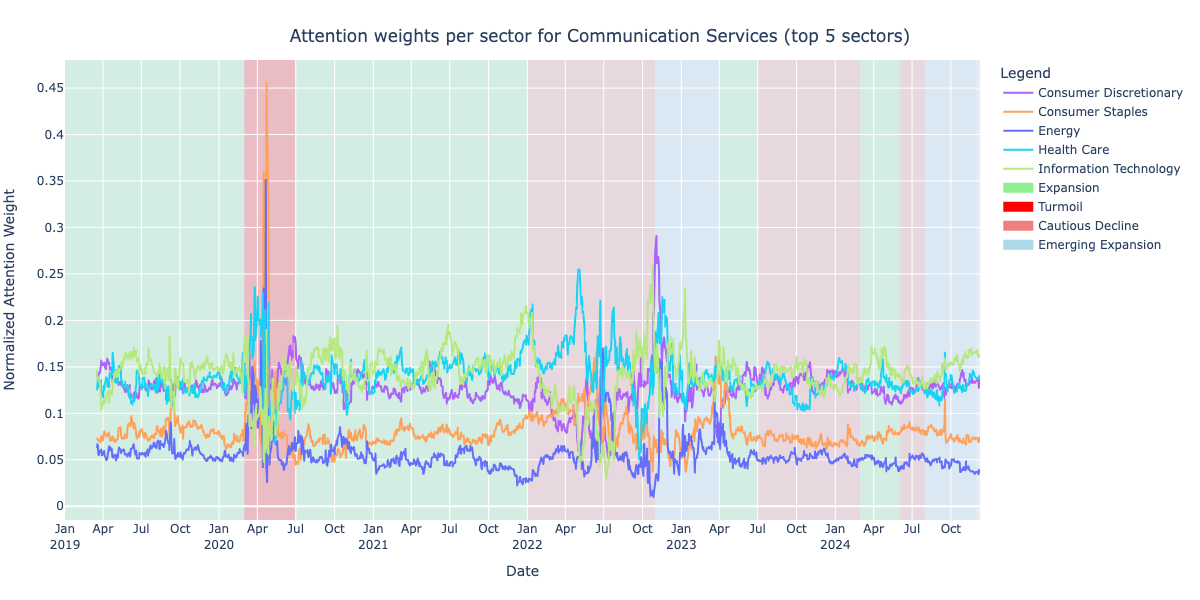} 
    \caption{Attention weights learned by the THGNN model for stocks in the Communication Services sector. Each data point shows the average attention a Communication Services stock gives to other sectors at a specific time, highlighting which sectors are the most important in forming predictions.}
    \label{fig:Communication Services} 
\end{figure}

\begin{figure}[H] 
    \centering
    \includegraphics[width=1.0\textwidth]{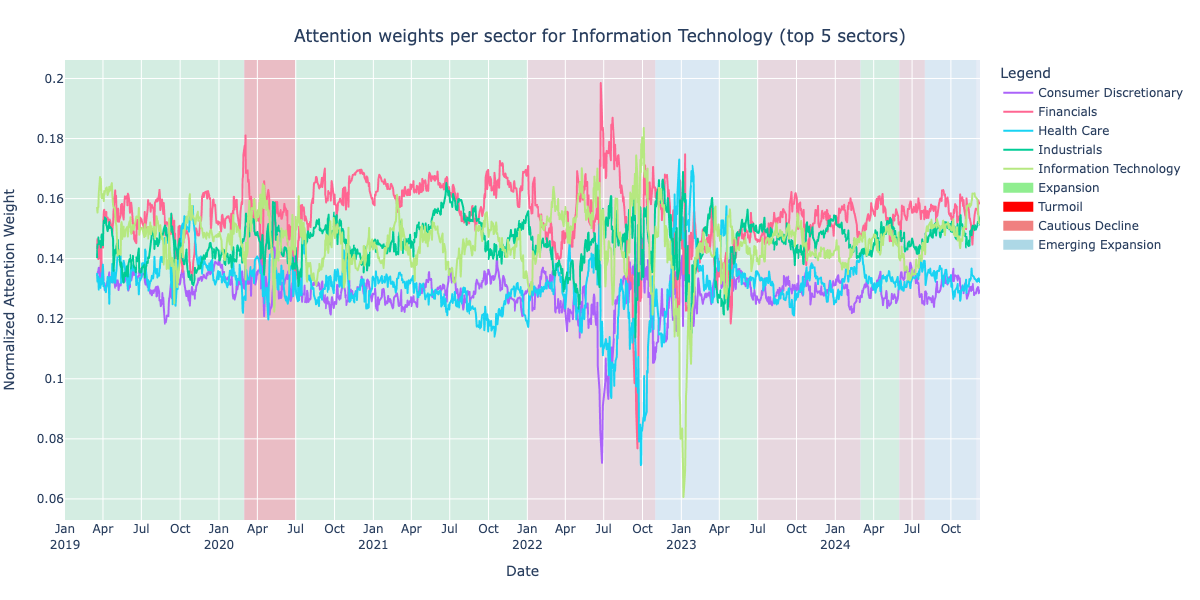} 
    \caption{Attention weights learned by the THGNN model for stocks in the Information Technology sector. Each data point shows the average attention an Information Technology stock gives to other sectors at a specific time, highlighting which sectors are the most important in forming predictions.}
    \label{fig:Information Technology} 
\end{figure}

\end{document}